\documentclass{article}

\usepackage{graphicx}
\usepackage{subcaption}

\usepackage{hyperref}
\usepackage{enumitem}

\usepackage[accepted,nohyperref]{mlsys2023}

\usepackage{amsmath}
\usepackage{amssymb}
\usepackage{bm}
\usepackage[capitalise,nameinlink,noabbrev]{cleveref}
\usepackage{makecell}
\usepackage{pifont}
\usepackage{multirow}

\graphicspath{{figures/}}

\newif\ifblind
\blindfalse

\ifblind
\newcommand{\tutel}[1]{\textsc{Flex}}
\else
\newcommand{\tutel}[1]{\textsc{Tutel}}
\fi
\newcommand{\AtoA}[1]{All-to-All}

\newcommand{\red}[1]{\textcolor{red}{#1}}

\newcommand{\blue}[1]{#1}

\newcommand*\circled[1]{\textcircled{\raisebox{-0.9pt}{#1}}}

\newcommand\algorithmicprocedure{\textbf{procedure}}
\newcommand{\algorithmicendprocedure}{\algorithmicend\ \algorithmicprocedure}
\makeatletter
\newcommand\PROCEDURE[3][default]{%
  \ALC@it
  \algorithmicprocedure\ \textsc{#2}(#3)%
  \ALC@com{#1}%
  \begin{ALC@prc}%
}
\newcommand\ENDPROCEDURE{%
  \end{ALC@prc}%
  \ifthenelse{\boolean{ALC@noend}}{}{%
    \ALC@it\algorithmicendprocedure
  }%
}
\newenvironment{ALC@prc}{\begin{ALC@g}}{\end{ALC@g}}
\makeatother

\mlsystitlerunning{\tutel{}: Adaptive Mixture-of-Experts at Scale}

\begin{document}

\twocolumn[
\mlsystitle{\tutel{}: Adaptive Mixture-of-Experts at Scale}



\mlsyssetsymbol{equal}{*}

\begin{mlsysauthorlist}
\mlsysauthor{Changho Hwang}{equal,msr}
\mlsysauthor{Wei Cui}{equal,msr}
\mlsysauthor{Yifan Xiong}{equal,msr}
\mlsysauthor{Ziyue Yang}{equal,msr}
\mlsysauthor{Ze Liu}{msr}
\mlsysauthor{Han Hu}{msr}
\mlsysauthor{Zilong Wang}{ms}
\mlsysauthor{Rafael Salas}{ms}
\mlsysauthor{Jithin Jose}{ms}
\mlsysauthor{Prabhat Ram}{ms}
\mlsysauthor{Joe Chau}{ms}
\mlsysauthor{Peng Cheng}{msr}
\mlsysauthor{Fan Yang}{msr}
\mlsysauthor{Mao Yang}{msr}
\mlsysauthor{Yongqiang Xiong}{msr}
\end{mlsysauthorlist}

\mlsysaffiliation{msr}{Microsoft Research}
\mlsysaffiliation{ms}{Microsoft}

\mlsyscorrespondingauthor{Yongqiang Xiong}{yqx@microsoft.com}

\mlsyskeywords{Machine Learning, MLSys}

\vskip 0.3in

\begin{abstract}


Sparsely-gated mixture-of-experts (MoE) has been widely adopted to scale deep learning models to trillion-plus parameters with fixed computational cost. The algorithmic performance of MoE relies on its token routing mechanism that forwards each input token to the right sub-models or \textit{experts}. While token routing dynamically determines the amount of expert workload at runtime, existing systems suffer inefficient computation due to their \textit{static execution}, namely static parallelism and pipelining, which does not adapt to the dynamic workload.

We present \tutel{}, a highly scalable stack design and implementation for MoE with dynamically adaptive parallelism and pipelining.
\tutel{} designs an identical layout for distributing MoE model parameters and input data, \blue{which can be leveraged by
switchable parallelism and dynamic pipelining methods without
mathematical inequivalence or tensor migration overhead.}
This enables adaptive parallelism/pipelining optimization at \textit{zero cost} during runtime.
Based on this key design, \tutel{} also implements various MoE acceleration techniques including Flexible \AtoA{}, two-dimensional hierarchical (2DH) \AtoA{}, fast encode/decode, etc. Aggregating all techniques, \tutel{} finally delivers
\textbf{4.96$\times$} and \textbf{5.75$\times$} speedup of a single MoE layer over 16 and 2,048 A100 GPUs, respectively, over the previous state-of-the-art.

Our evaluation shows that \tutel{} efficiently and effectively runs a real-world MoE-based model named SwinV2-MoE, built upon Swin Transformer V2, a state-of-the-art computer vision architecture. On efficiency, \tutel{} accelerates SwinV2-MoE, achieving up to $1.55\times$ and $2.11\times$ speedup in training and inference over Fairseq, respectively. On effectiveness, the SwinV2-MoE model achieves superior accuracy in both pre-training and down-stream computer vision tasks such as COCO object detection than the counterpart dense model, indicating the readiness of \tutel{} for end-to-end real-world model training and inference.

\end{abstract}
]



\printAffiliationsAndNotice{\mlsysEqualContribution} 

\newcommand{\speeduponA}{$3.11\times$}
\newcommand{\speeduponB}{$5.75\times$}
\newcommand{\speedupforfairseqeval}{$2.11\times$}
\newcommand{\speedupforfairseq}{$1.55\times$}

\section{Introduction}

\blue{In recent years,}
the community has found that enrolling more model parameters is one of the most straight-forward but less sophisticated way to improve the performance of deep learning (DL) algorithms~\citep{scaling-nlp}.
However, model capacity is often limited by computing resource and energy cost~\citep{nlp-cost}.
To tackle this, sparsely-gated Mixture-of-Experts (MoE)~\citep{sparsely-gated-moe} introduces a \textit{sparse} architecture by employing multiple parallel sub-models called \textit{experts}, where each input is only forwarded to a few experts based on an intelligent gating function.
Unlike dense layers, this method scales the model capacity up at only sublinearly increasing computational cost.
Nowadays, MoE is one of the most popular approaches demonstrated to scale DNNs to trillion-plus parameters~\citep{switch-transformer}, paving the way for models capable of learning even more information.

\begin{figure}[t]
    \centering
    \includegraphics[width=\linewidth]{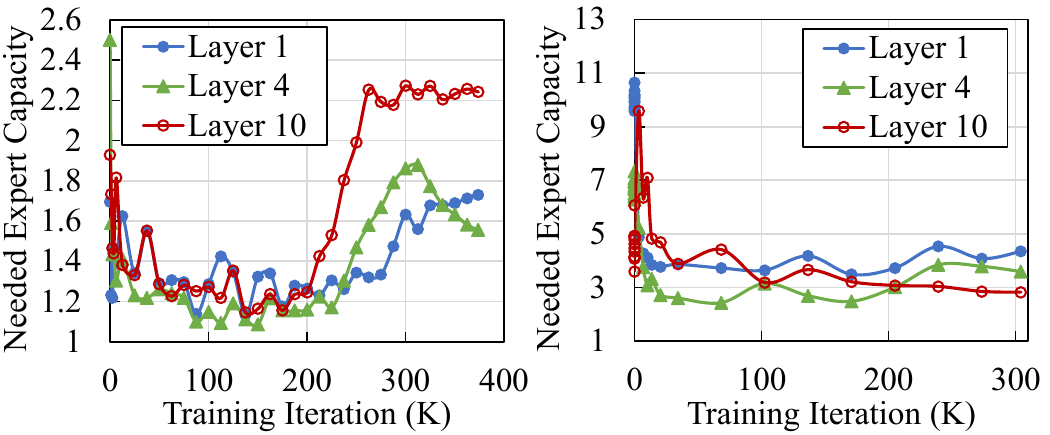}
    \caption{Dynamically changing workload of MoE layers during an end-to-end training of the MoE version of Swin Transformer V2~\cite{swin,liu2021swinv2} thin-tiny (left) and base (right) models. The y-axis is the needed expert capacity at runtime, which indicates the amount of workload (see details in \cref{ssec:background}). For a neat view, only the 1st, 4th, and 10th layers are shown out of 10 total MoE layers in the model.}
    \label{fig:workload}
    \vspace{-1em}
\end{figure}

While MoE-based algorithms open up a huge scale-up/out opportunity, the \textbf{dynamic nature of MoE} introduces fundamental system-side challenges that have not been seen before in most of previous DL algorithms and systems.
To be specific, each MoE layer consists of a certain number of parallel experts that are distributed over accelerators (GPUs in this work), where each GPU dispatches each input data to several best-fit experts according to an intelligent gating function and get the corresponding outputs back to combine them.
This implies that the workload of experts is fundamentally uncertain -- it depends on input data and the gating function. Both of them change at every iteration in practice.
In our experiments (see \cref{fig:workload}), the workload changes up to $4.38\times$ in a single training and different layers have different workload.

Previous DL systems, including the latest MoE frameworks~\citep{gshard,fairseq,deepspeed-moe,he2022fastermoe}, are mostly based on static runtime execution that does not fit dynamic MoE characteristics. The major pitfall comes from that experts often fail to leverage the best-performing parallelism because the optimal one differs depending on the dynamic workload.
It is non-trivial to dynamically adjust parallelism at runtime as it typically incurs a large redistribution overhead or GPU memory consumption in existing systems.
Other approaches such as \textit{load balancing loss}~\cite{switch-transformer} try to tackle this issue by manipulating the MoE algorithm, but it often harms model accuracy in our experiments (see \cref{ssec:background}).

This paper presents \tutel{}, a system that thoroughly optimizes MoE at any scale by adaptive methods specialized for dynamic MoE workload.
The key mechanism is \textit{adaptive parallelism switching} that dynamically switches the parallelism strategy at every iteration without any extra overhead of switching. Specifically, unlike existing systems that use different tensor layouts for different parallelism strategies, we leverage only a single distribution layout that covers all possibly optimal strategies. This frees the system from reformatting the input data or weights when we switch the parallelism strategy, hence zero-cost switching. Based on our communication cost analysis of all kinds of parallelism, we ensure that adaptive parallelism does not compromise the optimal parallelism strategy.

\tutel{} is a fully implemented framework for diverse MoE algorithms at scale. Over the adaptive parallelism switching, it delivers several optimization techniques for efficient and adaptive MoE, including adaptive pipelining, the 2-dimensional hierarchical (2DH) All-to-All algorithm, fast encode/decode with sparse computation on GPU, etc.
\ifblind
\else
\tutel{} has been open sourced on GitHub\footnote{\url{https://github.com/microsoft/tutel}} and already been integrated into Fairseq~\cite{fairseq} and DeepSpeed~\cite{deepspeed}. Our extensive experiments over Azure A100 clusters~\cite{ndmv4} show that with 128 GPUs, \tutel{} delivers up to \speeduponA \ of MoE-layer speedup, and \speedupforfairseq  / \speedupforfairseqeval\  speedup for end-to-end training / inference of a real-world model (SwinV2-MoE), compared to that of using the original Fairseq. For 2,048 GPUs, the MoE-layer speedup is further improved to \speeduponB.
\fi

Our key contributions are as follows:
\begin{itemize}[leftmargin=10pt]
  \setlength{\itemsep}{1pt}
  \setlength{\parskip}{0pt}
  \setlength{\parsep}{0pt}
    \item Provide detailed analysis on the dynamic nature of MoE and following challenges in existing frameworks.
    \item Propose adaptive parallelism switching that efficiently handles dynamic workload of MoE, which achieves
    \blue{$1.35\times\sim14.57\times$}
    speedup of a single MoE layer.
    \item Aggregating all acceleration techniques, \tutel{} delivers
    speedup of MoE at any scale: $4.96\times$ and $5.75\times$ speedup of a single MoE layer over 16 and 2,048 A100 GPUs, respectively.
    \item \tutel{} has been used to implement and run the sparse MoE version of a state-of-the-art vision model, SwinV2-MoE, on real-world computer vision problems. It achieves up to \speedupforfairseq\ and \speedupforfairseqeval\ speedup for training and inference, respectively, compared to previous frameworks such as Fairseq. We also demonstrate superior accuracy of the sparse model than the counterpart dense model, indicating the readiness of \tutel{} in training real-world AI models.
\end{itemize}

\section{Background \& Motivation}

This section introduces the dynamic nature of Mixture-of-Experts and its inefficiency in large-scale training.

\subsection{Background \& Related Work}
\label{ssec:background}

\paragraph{Sparsely-gated Mixture-of-Experts (MoE).}

\begin{figure}[t]
    \centering
    \includegraphics[width=\linewidth]{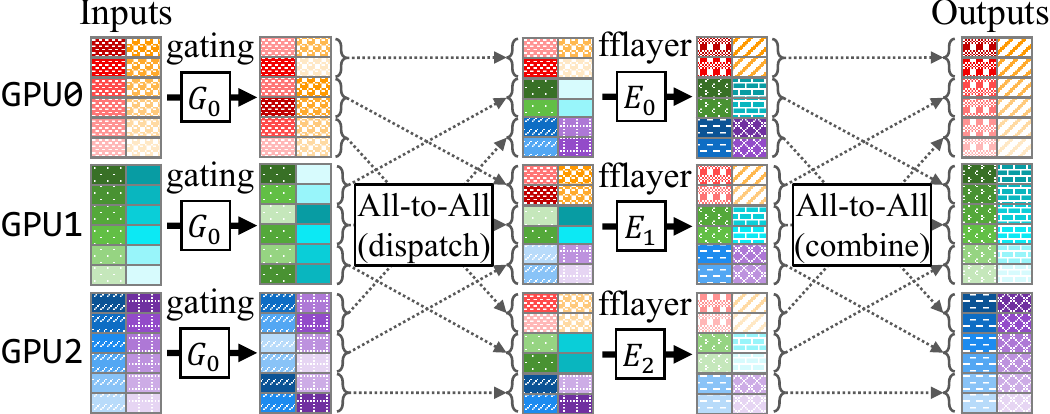}
    \caption{Example of an MoE layer across three GPUs, expert $E_i$ on GPU $i$. $G_0$ represents the gating function that is shared across all GPUs. Different colors or patterns indicate different samples (columns of inputs) and different gradients of color indicate different tokens within a sample (rows of inputs). This example shows two samples/batch, six tokens/sample, and evenly dispatched top-1 routing with capacity factor 1.0 -- see details in \cref{ssec:adaptive}.}
    \vspace{-1em}
    \label{fig:moe}
\end{figure}

MoE employs multiple \textit{expert} models, which deal with their own specialized sub-tasks respectively to solve the entire tasks together.
It is leveraged by large-scale distributed DNN models by putting a cross-GPU layer that partially exchanges hidden features from different GPUs~\cite{switch-transformer,m6,v-moe}.
\cref{fig:moe} illustrates an example.
First, it runs a \textit{gating function}~\cite{base-layers,hash-layers,m6-t} that determines the destination GPU of each input token\footnote{Each input sample is divided into one or more tokens, and the definition of a token depends on the model's algorithm and tasks.} in the following all-to-all collective communication (\AtoA{}).
After the \AtoA{} (called \textit{dispatch}), each GPU runs their own expert, which is a feed-forward network layer (fflayer), and then conducts the second \AtoA{} (called \textit{combine}) that sends the corresponding output of each token to the GPU where the token is from.
Details of the gating function and the fflayer defer depending on the model algorithm.

\paragraph{MoE as the Key to Exa-scale Deep Learning.}
MoE is differentiated from existing scale-up approaches for DNNs (i.e., increasing the depth or width of DNNs) in terms of its high cost-efficiency. Specifically, enrolling more model parameters (experts) in MoE layers does not increase the computational cost per token. Nowadays, MoE is considered as a key technology for hyper-scale DL with its state-of-the-art results shown in previous works~\cite{switch-transformer,v-moe,gshard,glam}. Currently, many state-of-the-art frameworks (e.g., DeepSpeed~\cite{deepspeed}, Fairseq~\cite{fairseq}, etc.) have already supported MoE.

\paragraph{Dynamic Workload of MoE.}
\blue{The root cause of dynamic workload of MoE
comes from
its token routing mechanism. Specifically, MoE layers dynamically route each token to multiple experts, where the distribution of tokens is often uneven across experts.}
This makes the workload of each expert dynamically change at every iteration as shown in \cref{fig:workload}.
\blue{\textit{Expert Capacity} is a common practice to indicate the workload of each expert, which is the number of tokens that an expert receives to deal with.}
Expert capacity depends on the number of tokens per batch $T$, the number of global experts $E$, top-$k$ routing ($1 \le k \le E$), and the capacity factor $f$ ($f\ge1$) as follows:
\begin{align}\label{fig:capacity}
Expert\ Capacity = k \cdot f \cdot \frac{T}{E}.
\end{align}
\blue{$f=1$ is
the minimum value indicating
the most even token distribution.
A larger $f$ value indicates
more imbalanced token routing,
which means that
an expert has to deal with more tokens.
}

\blue{Most existing MoE frameworks~\cite{fairseq,gshard,deepspeed-moe,alpa} simply set $f$ to a static upper bound of capacity factor $f_{upper}$ (i.e., $f=f_{upper}$) so that
different iterations always perform a static amount of computation.}
\blue{However,
static computation
based on $f_{upper}$ not only introduces unnecessary computations
but also may drop excessive tokens from training
if $f_{upper}$ is not set to a sufficiently large value, which
potentially impacts the model accuracy.
}
\blue{To tackle this, throughout this paper,
we consider a system (like \tutel{}) that supports
MoE training using the minimum required $f$
that incurs neither unneeded computation nor dropped tokens, as using $f=f_{upper}$ does.
Based on this mechanism, we explore further optimization opportunities
while $f$ varying across training steps.
}

\begin{table}[t]
\center\small
\begin{tabular}{c|cccccc}
\Xhline{1.0pt}
LB Loss Weight  & 0.001 & \textbf{0.01} & \textbf{0.1} & \textbf{1.0}\\ \hline
Acc@1 (\%)      & 37.32 & \textbf{37.78} & \textbf{37.16} & \textbf{34.71}\\\Xhline{1.0pt}
\end{tabular}
\caption{Harsh load balancing harms MoE model accuracy. \blue{Bold numbers highlight accuracy degradation with large LB loss weights.} All experiments are carried on ImageNet-22K image classification and the top-1 accuracy of SwinV2-S model is reported. Hyper-parameters: 32 experts, top-1 routing, capacity factor $f$=infinity.}
\label{table:lb_loss}
\vspace{-0.2em}
\end{table}

\paragraph{MoE Frameworks.}
While GShard~\cite{gshard} provides a computation logic that ensures algorithmic correctness of MoE, several popular MoE frameworks~\cite{fairseq,deepspeed-moe} follow the same logic but perform poorly on a large scale. Fast/\blue{FasterMoE}~\cite{he2022fastermoe} proposes different gating algorithms that are not computationally equivalent with GShard. Furthermore, it proposes \textit{shadow expert} and \textit{smart schedule} that 
\blue{deliver only conditional benefits when imbalanced token distribution persists for a long time, while may harm throughput otherwise.}
\blue{On the other hand,} \tutel{} pursues keeping the same computation logic as GShard and \blue{achieving} a deterministic gain
\blue{over any environments in general,}
which adapts MoE frameworks to exa-scale without harming algorithmic results.

\paragraph{Load Balancing Loss.}
Load balancing (LB) loss regulates MoE layer training by encouraging gating functions to balance workload of experts~\cite{sparsely-gated-moe,switch-transformer}.
LB loss can contribute to low and stable MoE workload as capacity factor $f$ typically decreases when the token distribution is even (as mentioned in the previous paragraph).
However, LB loss is typically insufficient to tackle the dynamic workload of MoE because giving a large weight on the LB loss often harms model accuracy.
Specifically, a proper weight on the LB loss may help model accuracy by guiding gating functions to enroll more diverse expert parameters during training, but a too large weight may harm the optimization objectives of the final task, as well as lead to failure of forwarding tokens to their knowledgeable experts.
\cref{table:lb_loss} shows that our experiments with large LB loss weights harm model accuracy.
Additionally, to our empirical findings, LB loss does not always result in more balanced workload across experts. For example, our experiments in \cref{fig:workload} use LB loss that help achieve the best accuracy, but it still shows dynamically changing workload.
In this paper, we only consider system-side solutions that are generally applied regardless of the LB loss.

\subsection{Static Parallelism}
\label{ssec:adaptive}


Under the dynamic nature of MoE layers, it becomes challenging if we would like to accelerate one expert with multiple GPUs for higher throughput. Previous research has proven that employing more experts typically gains only fast diminishing incremental benefits with many experts ($>256$)~\cite{deepspeed-moe,clark22,switch-transformer}. Therefore, in large-scale training, MoE layers typically employ relatively small number of experts compared with the number of GPUs and multiple GPUs are assigned to one expert for higher throughput.

We consider three different parallelism methods that have been adopted for MoE in prior works~\cite{switch-transformer}: expert parallelism (EP, distribute experts), data parallelism (DP, distribute input data), and model parallelism (MP, split and distribute a single expert). EP, DP, and MP can be used at the same time with each others.


\begin{figure}[t]
    \centering
    \includegraphics[width=\linewidth]{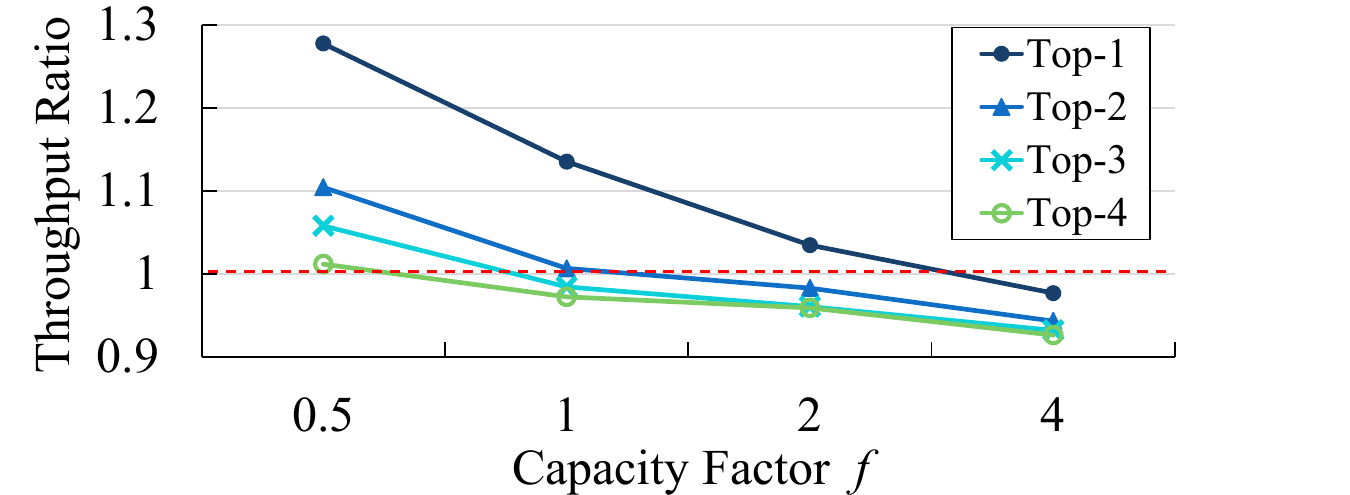}
    \caption{Runtime preferences of two different parallelism methods. The Y-axis measures the throughput ratio of EP+MP to EP+DP. It compares their throughput under varying capacity factor $f$ (i.e., varying amount of workload) and different top-$k$ configurations, where $>1.0$ implies that EP+MP outperforms EP+DP, and vice versa. Model settings: fflayer hidden size 16K, fflayer channel size 2048, and batch size 4.}
    \label{fig:pref-parallism}
\end{figure}

\begin{figure}[t]
    \centering
    \includegraphics[width=\linewidth]{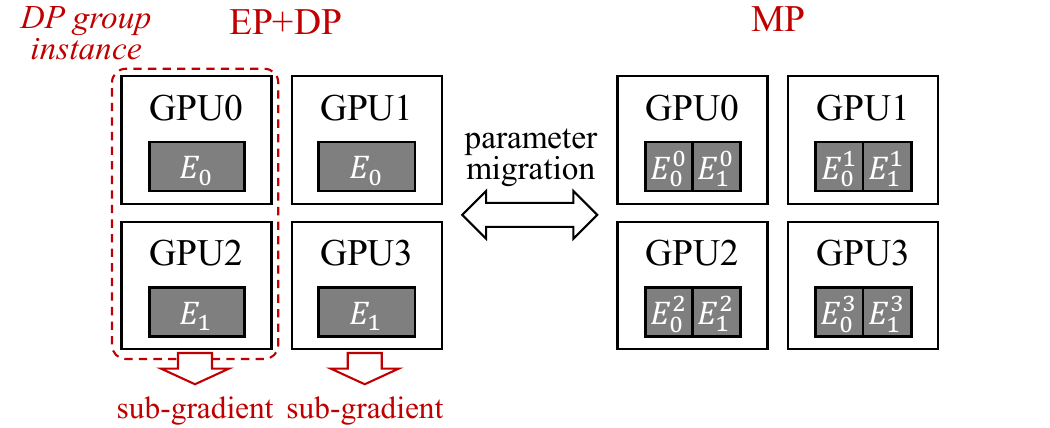}
    \caption{\blue{Parameter migration
    due to} \blue{switching parallelism between conventional EP+DP and MP}. $E_i^p$ refers to $p$-th slice (in the model-parallel manner) of $i$-th expert (no $p$ means not sliced). EP+DP replicates each expert on two GPUs each, and MP slices each expert across four GPUs respectively.
    }
    \label{fig:trivial-parallel}
\end{figure}

According to our experiments, statically adopting a certain parallelism method does not always work efficiently under dynamic workload.
For example, \cref{fig:pref-parallism} compares performance of two different parallelism methods, EP+DP and EP+MP.
As shown in the figure, the best parallelism method depends on the workload, which has 7.39\%-27.76\% performance gap between these two parallelisms.

Unfortunately, switching between different parallelism methods during runtime would incur a substantial overhead. Specifically, in existing work, an on-going training based on a certain parallelism (e.g., data-parallel) is not designed to be compatible with another parallelism (e.g., model-parallel) because they have different requirements on data split, weight split, managing momentum of parameter gradients, and even the framework interfaces to launch the training.
Furthermore, parameter migration is another costly overhead that would be incurred when we change the parallelism, as illustrated in \cref{fig:trivial-parallel}.
These are why parallelism switching is hardly used in existing systems.
 
\subsection{Static Pipelining}
\label{ssec:static-pipelining}

\begin{table}[t]
\center\small
\begin{tabular}{cccc}
\Xhline{1.0pt}
Number of GPUs & 16 & 64 & 256 \\ \hline\hline
MoE overhead (ms)   & 560.9 & 698.9 & 866.4  \\
Computation overhead (ms)  & 371.8   & 375.1   & 386.3  \\
\AtoA{} overhead (ms)  & 189.1   & 323.8   & 491.3  \\
\AtoA{} overhead ratio  & 33.7\%   & 46.3\% & 56.7\% \\ \hline\hline
Potential overhead saving & 33.7\%   & 46.3\% & 43.3\% \\ \hline
\textbf{Potential speedup} & \textbf{1.51$\times$} & \textbf{1.86$\times$} & \textbf{1.76$\times$} \\\Xhline{1.0pt}
\end{tabular}
\caption{Ratio of \AtoA{} overhead and potential speedup by fully overlapping \AtoA{} and computation in a typical MoE setting. Model settings: fflayer hidden size 4K, fflayer channel size 4K, 2 experts per GPU, 64K tokens per iteration.}
\label{table:background_overlapping_a2a_ratio}
\end{table}

MoE layers shown in~\cref{fig:moe} often under-utilize GPUs as they run \AtoA{} and fflayer in sequence to dispatch and combine. As \AtoA{} mostly consists of inter-GPU data copies that are not compute-intensive, we can better utilize computational power of GPUs by pipelining it with fflayer that runs numeric computation.
\cref{table:background_overlapping_a2a_ratio} shows up to $1.86\times$ potential speedup by overlapping \AtoA{} and fflayer computation.

\begin{figure}[t]
    \centering
    \includegraphics[width=\linewidth]{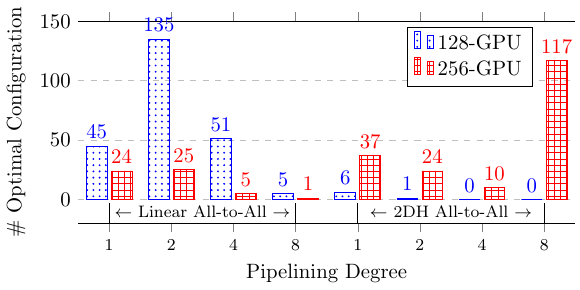}
    \caption{The distribution of optimal pipeline strategies for various MoE workload configurations. Each column indicates the number of configurations that perform best with the strategy described on X-axis. Details of workload configurations are the same as described in \cref{sec:eval_overlapping}.
    }
    \label{fig:pipeline}
\end{figure}

However, we observe that the static pipelining strategy for dispatch and combine, namely static \AtoA{} algorithm and pipelining degree, are inefficient to handle the dynamic workload. As illustrated in~\cref{fig:pipeline}, depending on different MoE settings and scales, the corresponding optimal pipelining strategy consists of various \AtoA{} algorithms (\textit{Linear} or \textit{2DH}\footnote{While Linear \AtoA{} lets all GPUs directly communicate with each others, 2DH (2-Dimensional Hierarchical) \AtoA{} adopts a hierarchical algorithm that conducts intra-node communication in a separate earlier stage. 2DH tends to outperform Linear on a larger scale, and vice versa. See details in \cref{apx:2dh}.}) and pipelining degrees.
This means that a single static strategy cannot always achieve the optimal performance in different MoE settings and scales, and dynamic pipelining strategy is necessary at runtime to adapt to varying settings.

To make things worse, the interference between computation and communication makes it difficult to find the optimal pipelining strategy if we only consider each single aspect separately.
This is because the slowdown from running NCCL kernels concurrently with computation kernels on the same GPU is difficult to estimate.
To our extensive experiments, even when two different \AtoA{} algorithms have similar throughputs, their throughputs often differ a lot when the same concurrent computation kernel is introduced, and either algorithm may outperform another one case-by-case.
This implies that the dynamic adjustment should be done jointly with both computation and communication
\blue{for the optimal overall throughput.}

\begin{table}
\centering\small
\begin{tabular}{c|c}\Xhline{1.0pt}
    Symbol & Description \\\hline\hline
    $W$ & World size used for \AtoA{} exchange \\\hline
    \blue{$D$} & fflayer channel size for each sample \\\hline
    \blue{$H$} & fflayer hidden size for each sample \\\hline
    \blue{$E_g$} & Number of local experts per GPU \\\hline
    $E$ & Number of global experts \\\hline
    \blue{$C_g$} & \blue{Token capacity per GPU}
    \\\hline
    $C$ & \blue{The total token capacity across GPUs} 
    \\\hline
    \blue{$P$} &
    \blue{The total parameters of all experts}
    \\\hline
    $f$ & The capacity factor used in \cref{fig:capacity}\\\hline 
\end{tabular}
\caption{Description of symbols.}
\label{tab:symbols}
\end{table}

\section{Adaptive MoE with \tutel{}}


\tutel{}, a full-stack MoE system, supports a complete MoE layer with adaptive optimizations. As all optimizations are transparent to DNN model developers, \tutel{} would not change the interface of DL frameworks and it can easily be integrated with other frameworks.
In the following subsections, we describe how \tutel{} tackles the aforementioned problems in detail.

\begin{table*}
\centering
\begin{tabular}{l|l|c|l} 
\Xhline{1.0pt}
Parallelism Method & Communication Complexity                                        & Limitation & Comment            \\ 
\hline\hline
\circled{1} DP       & \blue{$\mathcal{O}(P)$}       & -          & Possibly optimal   \\\hline
\circled{2} MP       & $\mathcal{O}(C_g \cdot W)$           & -          & No better than \circled{6}    \\\hline
\circled{3} EP       & $\mathcal{O}(C_g)$                   & $E/W \ge 1$    & No better than \circled{6}    \\\hline
\circled{4} DP+MP    & $\mathcal{O}(C_g \cdot r + \blue{P}/r)$     & $1 \le r \le W$  & No better than  \circled{7} for any $r$    \\\hline
\circled{5} EP+DP    & $\mathcal{O}(C_g + \blue{P/E})$        & -          & A special case of \blue{$r=1$ in \circled{7}} \\\hline
\circled{6} EP+MP    & $\mathcal{O}(C_g \cdot max\{1,W/E\})$   & -          & A special case of \blue{$r=W/E$ in \circled{7}} \\\hline
\circled{7} EP+DP+MP & \begin{tabular}[c]{@{}l@{}}$\mathcal{O}(C_g \cdot W/E)$~ -- if $r \ge W/E$\\$\mathcal{O}(C_g \cdot r + \blue{P/E} /r)$~ -- if $1 \le r < W/E$~\end{tabular} & -          & Possibly optimal   \\
\Xhline{1.0pt}
\end{tabular}
\caption{Analysis on communication complexity of MoE parallelism.}
\label{tbl:complexity}
\end{table*}

\subsection{Adaptive Parallelism Switching}
\label{ssec:design_adaptive_parallelism_switching}

\subsubsection{What is the least subset that is deserved for Parallelism Switching?}

Given that EP, DP, and MP derive 7 different possible combinations of parallelism methods, an ad-hoc approach is to design one execution flow for each method and makes it switchable with all other methods. However, designing up to 7 execution flows is not necessary as the problem can be precisely simplified into a smaller but \textit{efficiency-equivalent} problem, as is highlighted in the subsection title. 

Our approach is analyzing complexity of all parallelism methods to narrow them down to the least subset that we need to design execution flows for.
Note that only communication complexity matters here because all GPUs conduct an identical computation, hence the same computational complexity, so the communication complexities directly determines the efficiency of one parallelism method against others. As shown in \cref{tbl:complexity}, we analyze communication complexities of all parallelism methods to remove those from our consideration if they are (1) not the optimal in any cases or (2) a special case of another method.
By a series of comparison (shown in the Comment column of \cref{tbl:complexity}), we draw a conclusion that the subset can include only DP and EP+DP+MP. Therefore, the following paragraphs design corresponding parallel structure focusing \blue{only} on DP and EP+DP+MP, which still guarantees to cover the optimal parallelism method regardless of model configurations.

\subsubsection{Execution Flow of \textit{Zero Cost} Switchable Parallelism}

As explained in \cref{ssec:adaptive}, the switchable parallelism should guarantee exactly the same data layout and execution flow of MoE training. We explain our design for DP and EP+DP+MP respectively as follows. Zero Cost means that switching parallelism is completely free, without introducing any overhead larger than $\mathcal{O}(1)$ from parameter/token migration.

\begin{figure}[t]
    \centering
    \includegraphics[width=\linewidth]{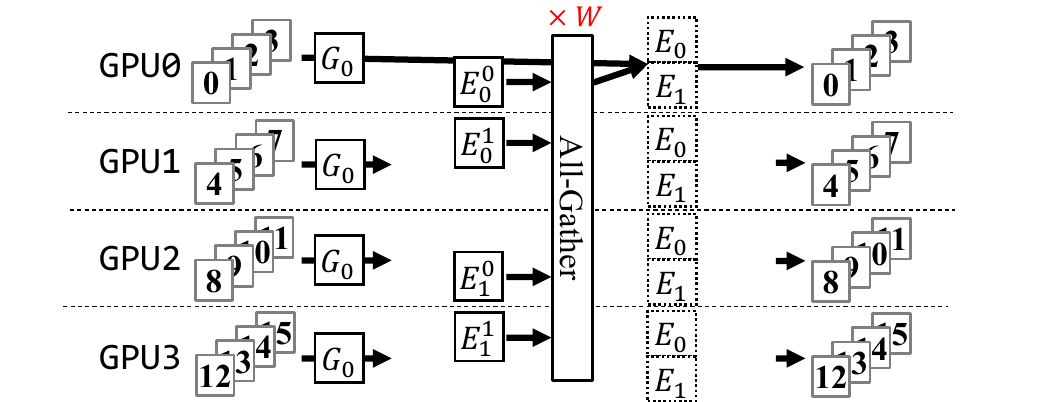}
    \caption{An example of DP execution flow in \tutel{}. All-gather is performed across all ($W$) GPUs.}
    \label{fig:exec-flow-dp}
    \vspace{-0.2em}
\end{figure}

\begin{figure}[t]
    \centering
    \includegraphics[width=\linewidth]{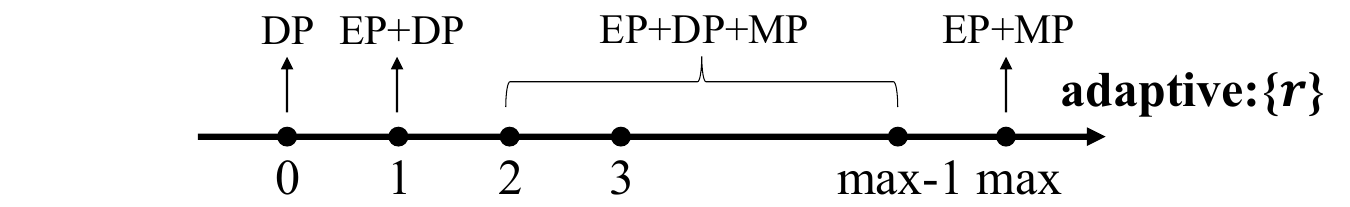}
    \caption{Specifying a parallelism method using \texttt{adaptive:r}, with \texttt{max} standing for the value $\lceil W/E \rceil$, and all $r$ values larger than this upper-bound are regarded the same as $\lceil W/E \rceil$.}
    \label{fig:adaptive_r}
    \vspace{-0.1in}
\end{figure}

\textbf{Switchable DP (\cref{fig:exec-flow-dp}):} It follows the conventional DP training that takes only local tokens as input, but
weight parameters following the ZeRO-DP Stage-3 Partitioning~\cite{zero} mechanism. Specifically, it lets each device to own a unique slice of weights, and performs one all-gather communication during the forward-pass and one reduce-scatter communication during the backward-pass, instead of the conventional training that performs one all-reduce communication during the backward-pass. Both ways are complexity-equivalent as a single all-reduce naturally consists of a reduce-scatter and an all-gather.
In \cref{fig:adaptive_r},
$r=0$ stands for the Switchable DP.

\begin{figure}[t]
    \centering
    \includegraphics[width=\linewidth]{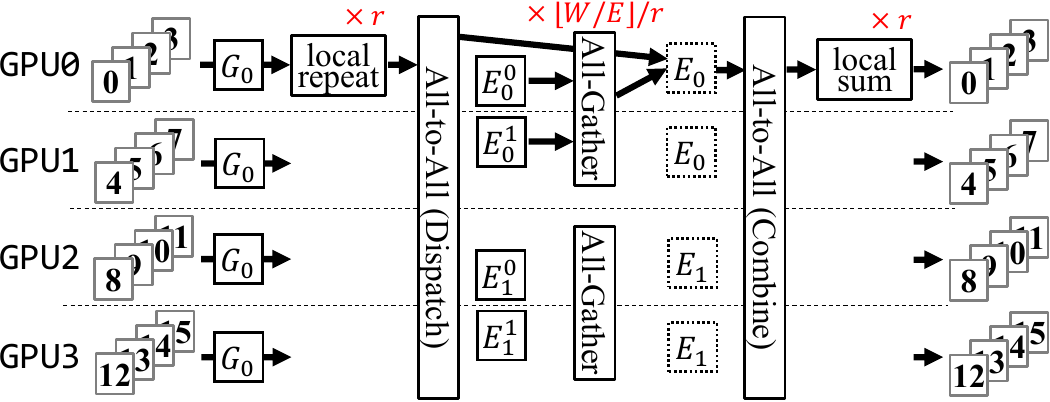}
    \caption{An example of EP+DP+MP execution flow in \tutel{}. Local repeat generates $r$ copies of gating function results, local sum reduces $r$ outputs from MoE combine, and all-gather is performed across $\lceil (W/E) /r\rceil$ GPUs.}
    \label{fig:exec-flow-epdpmp}
\end{figure}

\textbf{Switchable EP+DP+MP (\cref{fig:exec-flow-epdpmp}):} Out of the box, this parallelism method works the same as the Switchable DP -- they share the same format of reading inputs and slicing weights. Within the box, it not only ensures that the whole computation is mathematically equivalent to DP, but also ensures the required computation and network complexity
are within the expected complexity of EP+DP+MP as shown in \circled{7} of \cref{tbl:complexity}.
We define a control parameter $r$ that indicates to partition all GPUs into one or more groups with size $\lceil (W/E) /r \rceil$
each,
so that DP will be performed within each group and MP will be performed across different groups. Specifically, it repeats local tokens $r$ in the style of MP
at the beginning of execution flow, and finally performs a local sum symmetrically in the end. DP is only used to perform all-gather within a group of size $\lceil (W/E) /r \rceil$. Note that if $r$ increases and reaches $W/E$, the group size becomes 1, thus all-gather communication within each group is
\blue{optimized out}. This is why 
\blue{the case $r \ge W/E$ in \circled{7} eliminates an additional $\mathcal{O}(P / E / r)$.}
In \cref{fig:adaptive_r},
$r$ values from 1 to $W/E$ stands for the Switchable EP+DP+MP, though $r = 1$ and $r = \lceil W/E \rceil$ are two special cases that are exactly equivalent with EP+DP and EP+MP respectively.

\subsection{Adaptive Pipelining for Linear \& 2DH \AtoA{}}
\label{ssec:design_adaptive_piplining}

This section presents the design of adaptive pipelining.
As \AtoA{} communication latency substantially impacts on the optimal pipelining degree, our adaptive pipelining jointly optimizes both pipelining degree and \AtoA{} algorithms (Linear or 2DH) at the same time. While this section only explains how to partition input tokens for pipelining, the following \cref{ssec:search} describes how we jointly search for the optimal pipelining degree and the \AtoA{} communication algorithm.



\begin{figure}[t]
	\centering
    \includegraphics[width=\linewidth]{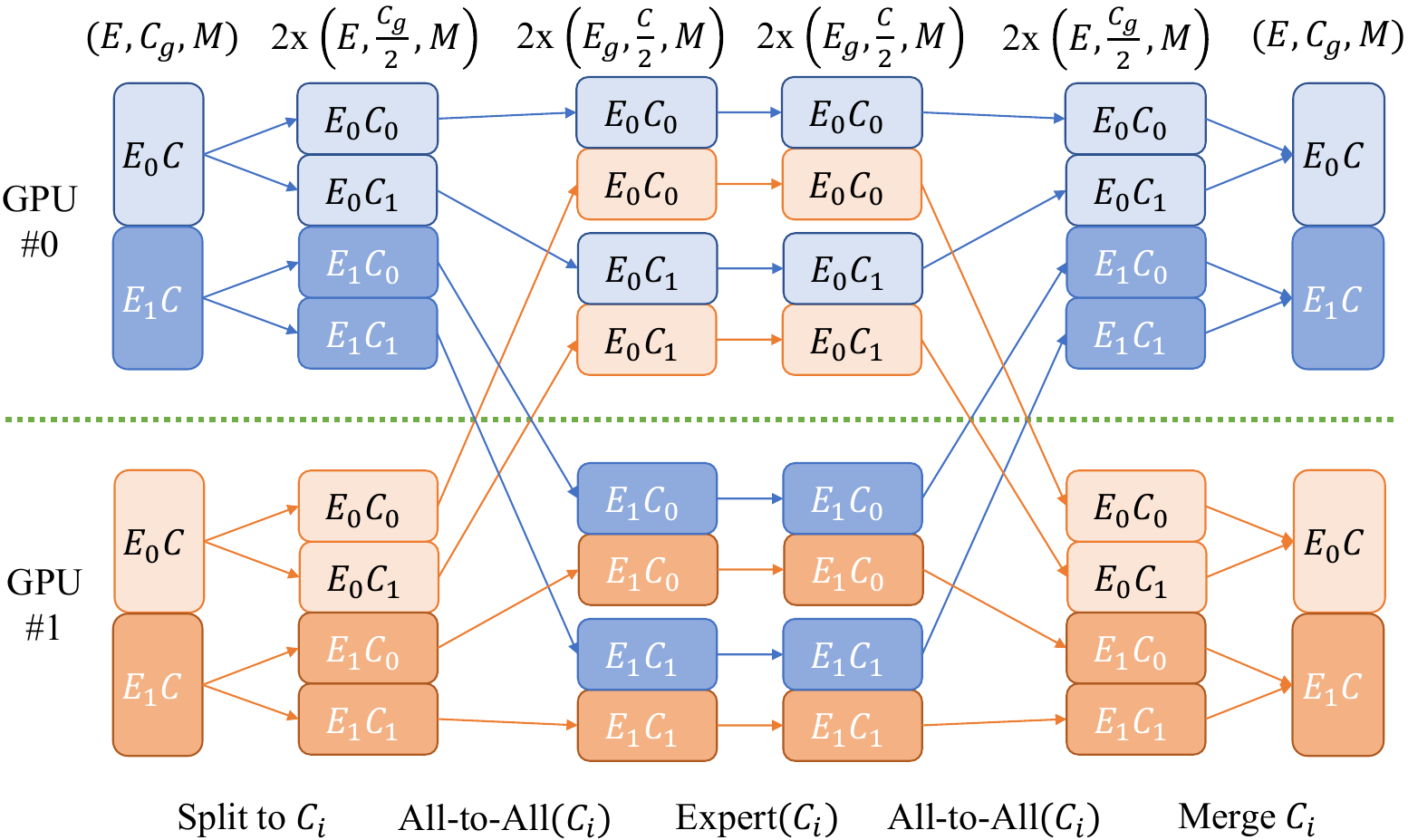}
	\caption{Overview of token partition on 2-expert-2-GPU for \AtoA{}-Expert multi-stream overlapping. $E_i$ means data is sent to i-th GPU and processed by $i$-th expert, and $C_i$ means data belongs to $i$-th partition of capacity dimension. \AtoA{} and expert operations of different capacity partitions can be overlapped.}
	\label{fig:overlap_a2a_overlap_overview}
\end{figure}

\paragraph{Token partition for multi-stream pipelining.}
Tokens need to be partitioned properly to enable the overlapping of flows on finer-grained data chunks, so that computation and communication can be submitted on separate GPU streams and run in parallel. Traditional partitioning like batch-splitting or pipeline-parallelism~\cite{gpipe} partitions all operations in the layer. This doesn't work in MoE because it amplifies the imbalance of MoE dispatch and destroys correctness for ML features like Batch Prioritized Routing~\cite{v-moe}. Instead, we propose to only partition the two \AtoA{}s and the expert in between instead of the whole MoE layer to avoid those shortcomings. \cref{fig:overlap_a2a_overlap_overview} gives 2-GPU example for data partition design in \AtoA{}-Expert overlapping.

In the forward pass, on each GPU, input of shape $(E, C_g, D)$ is split along dimension $C$ into two virtual partitions of shape $(E, C_g / 2, D)$. These two virtual partitions are marked with $C_0$ and $C_1$. 
After the splitting, each virtual partition $C_i$ is asynchronously sent to execute \AtoA{} operation in $i$'s order, on communication stream. \AtoA{} is customized to accept segregated data chunks as input and perform inline data shuffling, generating output of shape $(E_g, C/2, D)$.
Next, the two \AtoA{} outputs \blue{are} programmed to be sent to execute expert computation on computation stream once their previous corresponding \AtoA{} is completed, and the outputs of expert computation are again programmed to be sent to execute the second \AtoA{} on communication stream once previous corresponding expert computation is completed.
Finally, a barrier is set after the second \AtoA{}s, After the barrier, partitions are merged to generate final output of shape $(E, C_g, D)$.

The backward pass works in a similar way as the forward pass, except that the input becomes the gradients of the original output, the computation becomes the backward computation of the expert, and the output becomes the gradients of the original input.

Note that all partitioning and reshaping operations are done inline by customized operations, hence no extra data copy overhead compared with no-overlapping cases.

\subsection{Dictionary of Optimal Parallelism \& Pipelining}
\label{ssec:search}

\tutel{} manages a dictionary to memorize the optimal parallelism and pipelining setup of various different ranges of expert capacities. Specifically, we define the dictionary as a hash map:
$\lfloor c/R\rfloor \to \{r^*, d^*, a^*\},$
where $c$ is a capacity value of a certain iteration, $R$ is the window size that converges multiple adjacent $c$ values into the same key (default is 128), and $\{r^*, d^*, a^*\}$ is a tuple of the optimal setup (adaptive:$r$, pipelining degree, and \AtoA{} algorithm, respectively). To build up this dictionary beforehand, we need to find the optimal setup of each possible key ($\lfloor c/R\rfloor$) that only requires a few \blue{trials}, which is calculated as:
$$trials\ per\ key = (\log_{3/2}\lceil W/E \rceil + 2) \cdot \textit{4} \cdot \textit{2}.$$
$(\log_{3/2}\lceil W/E \rceil + 2)$ is the number of needed trials to search for $r^*$ via Ternary Search~\cite{ternary-search} because $r$ in range $[1,\lceil W/E \rceil - 1]$ determines a convex optimal distribution, plus two extra trials for $r=0$ and $r=\lceil W/E \rceil$. ``$\textit{4}$" is the number of needed \blue{trials} for $d^*$ as we limit the search space of the pipelining degree to $\{1,2,4,8\}$. To our practices, larger degrees than 8 hardly improve the overlapping between computation and communication, while significantly inflating \AtoA{} overhead. ``$\textit{2}$" refers to the number of \AtoA{} algorithms (Linear or 2DH).

\section{Implementation}
\subsection{Features}


    
    
    

\tutel{} provides more comprehensive support on MoE model training for different devices, data types and MoE-related features compared with other MoE frameworks, including DeepSpeed MoE, Fairseq MoE, and FastMoE.

\paragraph{\textit{Dynamic} Top-\textit{ANY} MoE Gating.}
To enable a variety of sparsity options for MoE training, \tutel{} supports top-ANY routing. The $k$ value can be customized per step as well to enable
dynamic sparsity updates,
which is useful when
different iterations of one MoE layer use their preferred top-$k$ settings instead of using the same $k$ value.
Users can leverage this feature to dynamically fine-tune sparsity of MoE layers.



\begin{figure}[t]
  \centering
  \includegraphics[width=\linewidth]{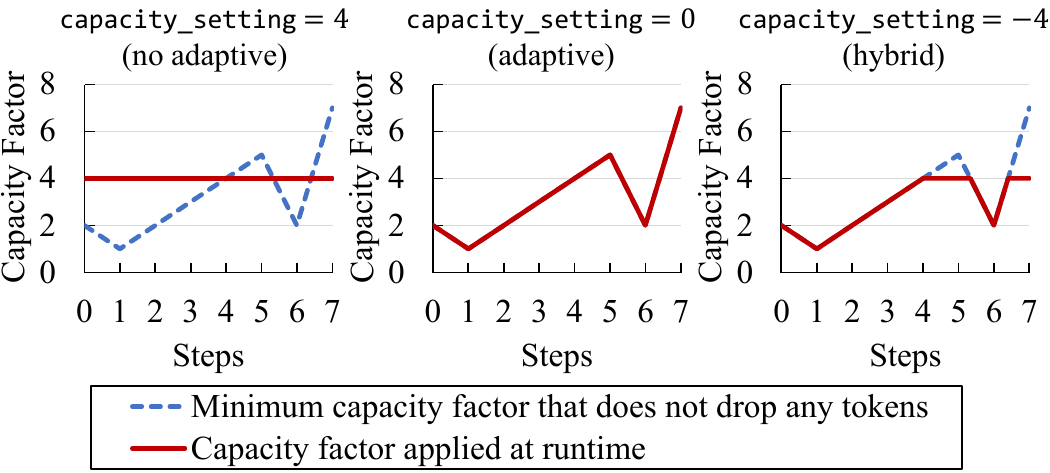}
  \caption{Examples of dynamic capacity factor adaptation when
  \blue{$capacity\_setting$}
  is given as $4$, $0$, and $-4$, respectively.}
  \label{fig:dynamic-cap-factor}
\end{figure}

\paragraph{\textit{Dynamic} Capacity Factor.}
To smartly control the capacity upper-bound under varying token imbalance, \tutel{} supports adjusting the capacity factor dynamically at every iterations. As illustrated in \cref{fig:dynamic-cap-factor}, the adjustment behavior is controlled by argument
\blue{$capacity\_setting=x$} passed to our MoE layer API. If $x$ is positive, the value is directly applied as the capacity factor of the MoE layer. If $x$ is zero, \tutel{} automatically adapts the capacity factor to the minimum value that does not drop any tokens at each iteration. If $x$ is negative, it works the same as when $x$ is zero except that $-x$ is set as the upper bound of capacity factor, i.e., any exceeding value will be adapted to $-x$.

\subsection{Optimizations}
\label{ssec:optimizations}

\begin{figure}[t]
  \centering
  \includegraphics[width=\linewidth]{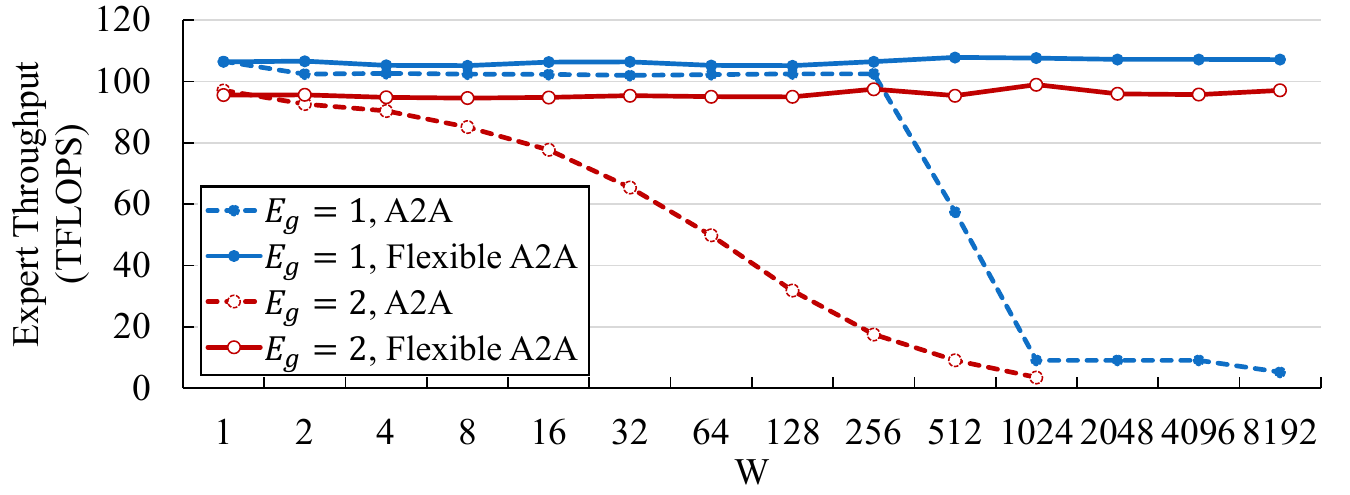}
  \vspace{-2em}
  \caption{Throughput for expert computation based on A2A (\AtoA{}) layout and Flexible A2A layout.}
  \vspace{-1em}
  \label{fig:layout}
\end{figure}

\paragraph{Flexible \AtoA{}.}
We propose an abstraction upon conventional MPI/NCCL \AtoA{} interfaces to ensure high computational throughput of MoE experts regardless of the scale, which is called \textit{Flexible \AtoA{}} in this context. Existing \AtoA{} transforms the tensor layout from $(E, C_g, D)$ into $(W, E_g, C_g, D)$ where $C_g$ relies on $W$, which affects the efficiency of the following matrix multiplication by experts. Instead, we transform the output layout into $(E_g, C, D)$ that ensures the same-shaped matrix multiplication at any scale ($W$).
\cref{fig:layout} compares the expert computation throughput between the conventional \AtoA{} and Flexible \AtoA{}.

\paragraph{Kernel Optimization: Fast Encode and Decode.}
According to GShard~\cite{gshard}, existing implementations for MoE dispatch and combine need multiple einsum and matrix multiplication operations. \tutel{} deeply optimizes this by using
SIMT-efficient
sparse operations,
which we call \textit{fast encode and decode}. It largely minimizes the latency of non-expert computations, as shown in \cref{fig:comp-breakdown}.
This optimization saves GPU memory as well, achieving $20\% \sim 90\%$ memory saving in most cases. See more details of fast encode and decode in \cref{apx:fast-enc-dec}.

\begin{table}[t]
\centering\small
\begin{tabular}{c|c|c}
    \Xhline{1.0pt}
    tokens/step & Fairseq MoE (GiB) & \tutel{} MoE (GiB) \\\Xhline{1.0pt}
    4,096 & 3.7 & 2.9 ~~ (-21.6\%) \\\hline
    8,192 & 6.2 & 3.2 ~~ (-48.4\%)  \\\hline
    16,384 & 16.3 & 4.0 ~~ (-75.5\%) \\\hline
    32,768 & 57.9 & 5.7 ~~ (-90.2\%) \\\Xhline{1.0pt}
\end{tabular}
\vspace{-1em}
\caption{GPU memory cost for single MoE layer. (Static Settings: $D = H = 4096$, top-k = 2, $E_g$ = 2)}
\vspace{-1em}
\label{tab:memory}
\end{table}

\section{Evaluation}
\label{sec:eval}

\paragraph{Testbed.}
If not specified, all experiments use Azure Standard\_ND96amsr\_A100\_v4 VMs~\cite{ndmv4} .
Each VM is equipped with $8\times$ NVIDIA A100 SXM 80GB GPUs and $8\times$ 200 Gbps HDR InfiniBand, backed by $96\times$ 2nd-generation AMD Epyc CPU cores and 1.9 TiB memory. GPUs are connected by 3rd-generation NVLink and NVSwitch within one VM, while different VMs are connected through 1,600 Gbps InfiniBand non-blocking network with adaptive routing.

\vspace{-1em}
\paragraph{Setup.}
For baseline, we use PyTorch 1.8.0 and Fairseq \texttt{moe} branch by default.
NCCL 2.10.3-1~\cite{nccl} and NCCL RDMA SHARP plugin~\cite{nccl-plugin} are used for communication when scaling out. We use up to 2,048 A100 GPUs (\blue{256} VMs) for experiments.

\vspace{-0.5em}
\subsection{Evaluation on Adaptive MoE with \tutel{}}
\label{ssec:micro-bench}

This section evaluates gains from adaptive computation using \tutel{}.
We compare the throughput of optimal parallelism / pipelining strategy
and study the gain from adaptivity of \tutel{}. For apples-to-apples comparison with existing frameworks, in \cref{ssec:eval-single-layer}, we compare \tutel{} with Fairseq MoE~\cite{fairseq} only using a specific parallelism method that is supported by both.

\subsubsection{Adaptive Parallelism Switching}
\label{ssec:eval-flexibility}

\begin{figure}[t]
  \centering
    \includegraphics[width=\linewidth]{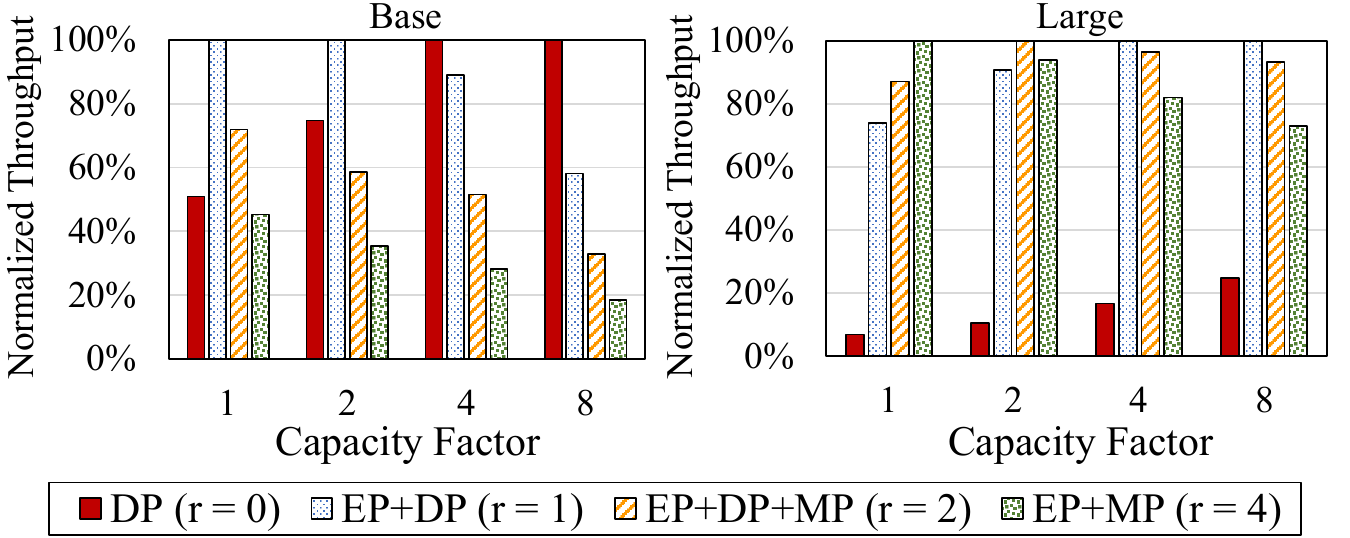}
    \vspace{-2em}
    \caption{Normalized throughputs of different
    \blue{capacity factors $f$}
    under \blue{Base (left) and Large (right)} MoE configurations.
    \blue{The figure shows that the optimal parallelism method differs depending on capacity factor $f$.}
    }
  \label{fig:parallel-eval}
\end{figure}

We evaluate adaptive parallelism switching with various MoE model settings using a single node.
\cref{fig:parallel-eval} compares normalized throughputs \blue{using different parallelism options where capacity factor $f$ varies from 1.0 to 8.0. We test two MoE configurations, Base ($samples/step = 4K$ and $H=2K$) and Large ($samples/step = 1K$ and $H=32K$), while other expert settings are shared ($E=16$, $D=2K$, and 64 total GPUs).}
As shown in the figure, the optimal parallelism method varies depending on the MoE expert configurations and capacity configurations.
For instance, DP ($r=0$) tends to be more favorable when the expert capacity is high, and as the capacity decreases, the tendency gradually changes to EP+DP ($r=1$) and then to EP+DP+MP ($r>1$).
\blue{In relatively smaller-scale MoE configurations, the optimal parallelism option typically stays in $r=0$ or $r=1$, while it dynamically changes across a wider range of $r$ values in larger-scale configurations.}
Such a variety evidences a \blue{substantial} chance of improvements with \tutel{}, which
leads
to different optimal parallelism methods according to
the dynamically changing \blue{$f$},
as explained in \cref{ssec:design_adaptive_parallelism_switching}.

\subsubsection{Adaptive Pipelining}
\label{sec:eval_overlapping}

\begin{table}[t]
\centering
\begin{subtable}[h]{0.45\textwidth}
\centering\small
\begin{tabular}{c|c|cccc}
\Xhline{1.0pt}
\multirow{2}{2em}{\centering\small GPU\\Num.} & \multirow{2}{3em}{\centering\small All2All\\Algo.} & \multicolumn{4}{c}{Pipelining Degree} \\
& & 1 & 2 & 4 & 8 \\
\Xhline{1.0pt}
\multirow{2}{2em}{\centering 16}
& Linear & 20\%  & 2\%  & 2\%  & 11\% \\
& 2DH & 101\%  & 98\%  & 100\%  & 106\% \\
\hline
\multirow{2}{2em}{\centering 32}
& Linear & 16\%  & 1\%  & 2\%  & 11\% \\
& 2DH & 45\%  & 43\%  & 44\%  & 51\% \\
\hline
\multirow{2}{2em}{\centering 64}
& Linear & 13\%  & 1\%  & 5\%  & 15\% \\
& 2DH & 28\%  & 25\%  & 27\%  & 34\% \\
\hline
\multirow{2}{2em}{\centering 128}
& Linear & 9\%  & 2\%  & 9\%  & 29\% \\
& 2DH & 16\%  & 16\%  & 19\%  & 26\% \\
\hline
\multirow{2}{2em}{\centering 256}
& Linear & 20\%  & 27\%  & 54\%  & 107\% \\
& 2DH & 12\%  & 20\%  & 34\%  & 11\% \\
\Xhline{1.0pt}
\end{tabular}
\caption{Adaptive pipelining improvement on average.}
\label{table:eval_adaptive_pipelining_average}
\end{subtable}
\begin{subtable}[h]{0.45\textwidth}
\centering\small
\begin{tabular}{c|c|cccc}
\Xhline{1.0pt}
\multirow{2}{2em}{\centering\small GPU\\Num.} & \multirow{2}{3em}{\centering\small All2All\\Algo.} & \multicolumn{4}{c}{Pipelining Degree} \\
& & 1 & 2 & 4 & 8 \\
\Xhline{1.0pt}
\multirow{2}{2em}{\centering 16}
& Linear & 60\%  & 32\%  & 50\%  & 176\% \\
& 2DH & 149\%  & 139\%  & 142\%  & 184\% \\
\hline
\multirow{2}{2em}{\centering 32}
& Linear & 60\% & 31\%  & 41\%  & 135\% \\
& 2DH & 89\%  & 75\%  & 59\%  & 148\% \\
\hline
\multirow{2}{2em}{\centering 64}
& Linear & 55\%  & 23\%  & 42\%  & 161\% \\
& 2DH & 70\%  & 54\%  & 41\%  & 109\% \\
\hline
\multirow{2}{2em}{\centering 128}
& Linear & 45\%  & 54\%  & 87\%  & 300\% \\
& 2DH & 52\%  & 37\%  & 35\%  & 107\% \\
\hline
\multirow{2}{2em}{\centering 256}
& Linear & 100\%  & 160\%  & 317\%  & 599\% \\
& 2DH & 73\%  & 139\%  & 193\%  & 182\% \\
\Xhline{1.0pt}
\end{tabular}
\caption{Adaptive pipelining improvement over the worst case.}
\vspace{-0.7em}
\label{table:eval_adaptive_pipelining_worse_case}
\end{subtable}
\caption{Adaptive pipelining improvements.}
\vspace{-1.5em}
\end{table}

\begin{figure}[t]
    \centering
    \includegraphics[width=\linewidth]{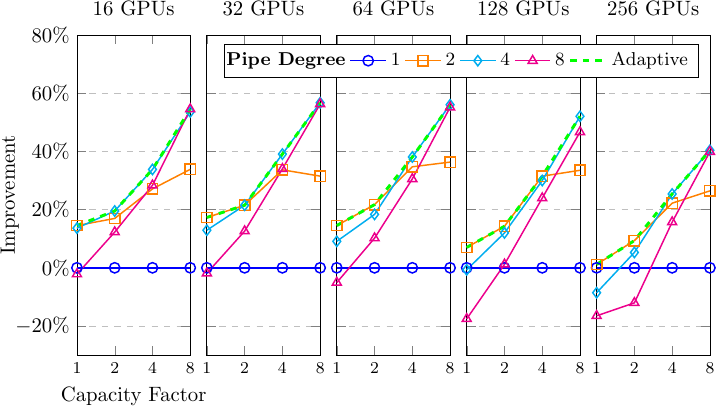}
    \caption{\blue{Improvement from adaptive pipelining depending on capacity factor $f$}. $D = 4096$, $H = 4096$, $E_g = 2$, and tokens/step = 4096.}
    \label{fig:eval_adaptive_pipelining_worse_case_dynamic_workload}
\end{figure}


We evaluate adaptive pipelining on 243 typical MoE model settings on different scale ($16\sim256$ GPUs). We test all combinations of MoE model configurations within: $E_g\blue{\in}\{0.5,1,2\}$, $D\blue{\in}\{1024,2048,4096\}$, $H\blue{\in}\{1024,2048,$ $4096\}$, and \mbox{tokens/step} $\blue{\in}\{4096,16384,65536\}$. For comparison, we also measure different static pipelining methods considering different degrees $\{1,2,4,8\}$ and different \AtoA{} algorithms (Linear or 2DH).

\cref{table:eval_adaptive_pipelining_average} shows average improvement on these 243 models. Compared with baseline solution (pipelining degree 1) and Linear \AtoA{}), adaptive piplining achieves $9\% \sim 101\%$ improvement in average. Compared with different static strategies, it also can achieve $1\% \sim 107\%$ improvement in average. Besides, adaptive piplining achieves significant improvement and avoids performance regression in the worst case, which shows $23\% \sim 599\%$ improvement in~\cref{table:eval_adaptive_pipelining_worse_case}.

We also evaluate the performance gain under different dynamic workloads on different scales. We use different capacity factors $f$ to emulate different workload patterns in different training iterations. As shown in~\cref{fig:eval_adaptive_pipelining_worse_case_dynamic_workload}, adaptive pipelining always chooses the best strategy, and it can achieve up to 39\% improvement with $f=4$ and up to 57\% improvement with $f=8$, compared with baseline (pipelining degree 1).

\begin{figure}[t]
  \centering
    \includegraphics[width=\linewidth]{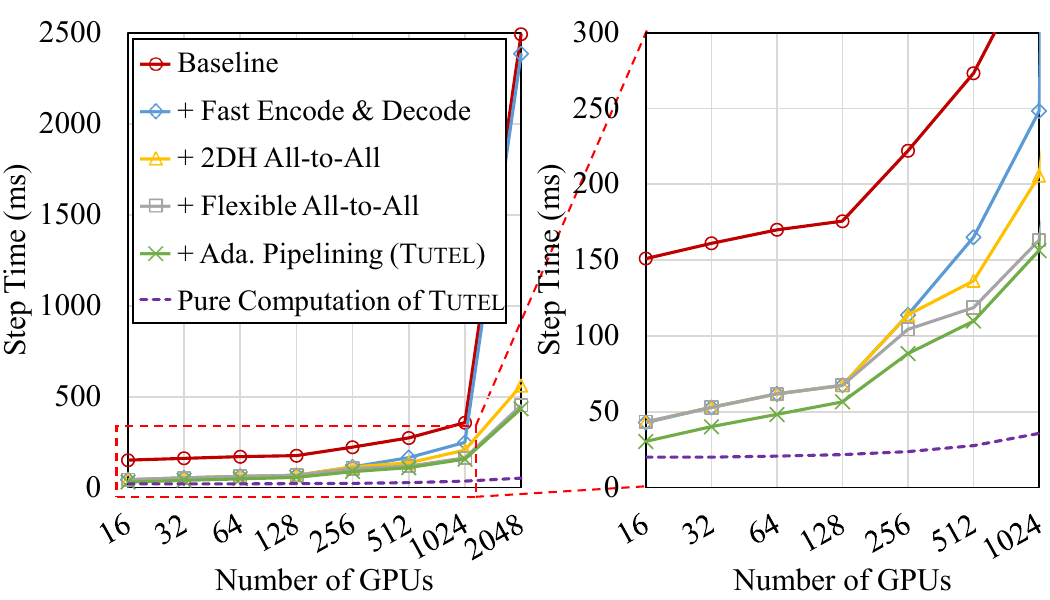}
    \vspace{-2em}
    \caption{Single MoE layer improvement breakdown. \blue{The baseline is a Fairseq / DeepSpeed MoE layer.}}
    \vspace{-1em}
  \label{fig:single-layer-breakdown}
\end{figure}

\begin{figure}[t]
    \centering
    \ifblind
\else
    \includegraphics[width=\linewidth]{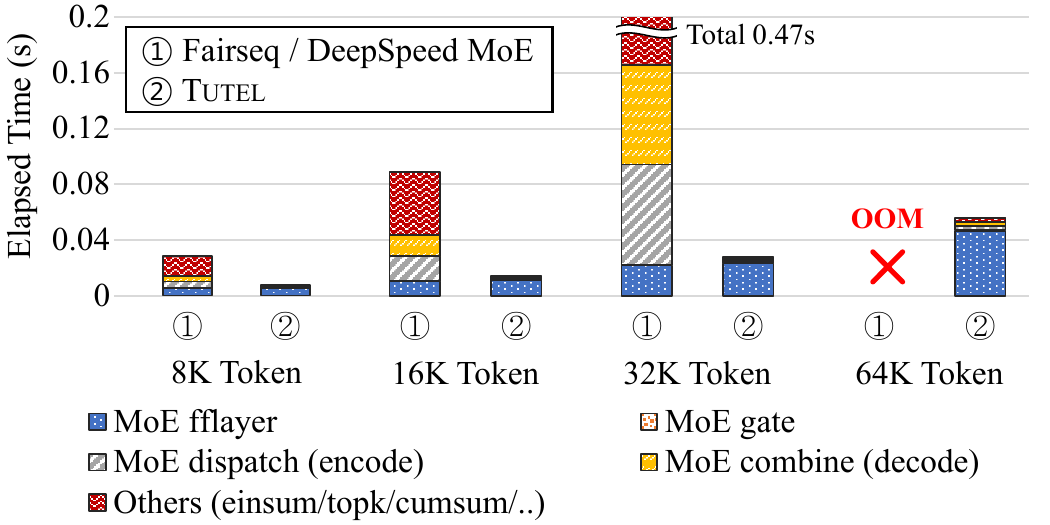}
    \fi
    \vspace{-2em}
    \caption{Kernel computation breakdown comparison between \tutel{} and \blue{Fairseq / DeepSpeed MoE.}}
    \label{fig:comp-breakdown}
    \vspace{-2em}
\end{figure}

\subsection{Single MoE Layer Scaling}
\label{ssec:eval-single-layer}

We evaluate the step time of single MoE layer when scaling out to 2,048 GPUs.
It uses tokens/step = 16384, $f$ = 1, $D$ = 2048, $H$ = 2048, $E_g$ = 2, top-$k$ = 2, adaptive:$r$ = 1.
We add \tutel{} features once at a time to study where the major gain is from, where Fairseq~\cite{fairseq} is used as the baseline.
Detailed experiments for each feature are provided in the following \cref{ssec:micro-bench}.

The following explains each curve in \cref{fig:single-layer-breakdown} in order.
\mbox{\circled{1} \blue{(red, circle)} Fairseq / DeepSpeed MoE} Baseline. \blue{Fairseq and DeepSpeed MoE perform the same as they use an equivalent MoE layer implementation.}
\circled{2} \blue{(blue, diamond)} \tutel{} Kernel (Fast Encode \& Decode in \cref{ssec:optimizations}) + Linear \AtoA{}.
\tutel{} kernel optimizations deliver a \blue{large} gain at a small scale ($3.52\times$ on 16 GPUs), while the gain becomes small at a large scale ($1.04\times$ on 2,048 GPUs). The detailed gains from using \tutel{} kernels over Fairseq are shown in~\cref{fig:comp-breakdown}.
\circled{3} \blue{(yellow, triangle)} \tutel{} Kernel + 2DH \AtoA{}.
2DH \AtoA{} delivers a significant gain on a large scale ($4.25\times$ on 2,048 GPUs).
\circled{4} \blue{(gray, square)} \tutel{} Kernel + 2DH \AtoA{} + Flexible \AtoA{}.
Flexible \AtoA{} delivers gains on large scales starting from 256 GPUs, e.g., $1.24\times$ on 2,048 GPUs compared with not using it.
\circled{5} \blue{(green, cross)} \tutel{} Kernel + 2DH \AtoA{} + Flexible \AtoA{} + Adaptive Pipelining Degree.
\circled{5} shows the mixture of gains from optimizing the pipelining degree together with Linear/2DH \AtoA{} algorithms, further achieving $1.43\times$ and $1.04\times$ on 16 and 2,048 GPUs, respectively. \circled{5} becomes less important on larger scales as the overhead of slicing tokens becomes more detrimental to \AtoA{} efficiency.
The breakdown does not include adaptive parallelism switching as it statically uses adaptive:$r$ = 1, not only because this parallelism is officially supported by Fairseq MoE while others are not, but also to ensure that the \AtoA{} communication size required by \tutel{} and Fairseq MoE are exactly the same, so as to fairly compare the improvement of \AtoA{}.

Compared with the baseline, \tutel{} finally delivers $\bm{4.96\times}$, \blue{$\bm{3.11\times}$,} and $\bm{5.75\times}$ speedup on 16 GPUs, \blue{128 GPUs,} and 2,048 GPUs, respectively. 
For computation-communication breakdown, \circled{6} \blue{(purple, dashed)} shows the pure computation overhead of the complete \tutel{} (excluding the portion overlapped with communication).
Note that the slight increase of computation overhead as we scale out is not from the system overhead but due to more theoretical computation required by the gating function for total $E_g \cdot W$ experts.

\subsection{Adoption to Real-world Problems: SwinV2-MoE}
\label{ssec:eval-swin-moe}


We introduce SwinV2-MoE to verify the correctness and performance of \tutel{} in end-to-end training and testing. SwinV2-MoE is an MoE version of Swin Transformer V2~\cite{swin, liu2021swinv2}, which is a state-of-the-art computer vision neural network architecture that is widely used in a large variety of computer vision problems. SwinV2-MoE is built from a dense Swin Transformer V2 model with every other feed-forward layer replaced by an MoE layer except for the first two network stages. The SwinV2-B model is adapted for experiments, and the default hyper-parameters are: $E=32$, top-$k=1$, \blue{and} $f=1.0$.

\subsubsection{Experiment Setup}

\paragraph{Pre-training and Down-stream Computer Vision Tasks.} We follow~\cite{swin} to use ImageNet-22K image classification datasets for model pre-training, which contains 14.2 million images and 22 thousand classes. In addition to evaluating the performance of the pre-training task (using a validation set with each class containing 10 randomly selected images), we also evaluated the models using 3 down-stream tasks: 1) ImageNet-1K fine-tuning accuracy. The pre-trained models are fine-tuned on ImageNet-1K training data and the top-1 accuracy on the validation set is reported; 2) ImageNet-1K 5-shot linear evaluation~\cite{v-moe}. 5 randomly selected training images are used to train a linear classifier, and the top-1 accuracy on the validation set is reported; 3) COCO object detection~\cite{lin2014microsoft}. The pre-trained models are fine-tuned on the COCO object detection training set using a cascade mask R-CNN framework~\cite{swin}, and box/mask AP on the validation set is reported.

\begin{table}[t]
\centering
\small
\addtolength{\tabcolsep}{-2.pt}
\begin{tabular}{ccccc}
    \Xhline{1.0pt}
    \#GPU & \makecell{Dense\\train / infer} & \makecell{Fairseq MoE\\train / infer} & \makecell{\tutel{} MoE\\train / infer} & \makecell{Speedup\\train / infer} \\
    \hline
    8 & 291 / 1198 & 240 / 507 & 274 / 1053 & 1.14$\times$ / 2.08$\times$ \\
    16 & 290 / 1198 & 173 / 473 & 253 / 943 & 1.46$\times$ / 1.99$\times$ \\
    32 & 288 / 1195 & 162 / 455 & 249 / 892 & 1.54$\times$ / 1.96$\times$ \\
    64 & 285 / 1187 & 159 / 429 & 234 / 835 & 1.47$\times$ / 1.95$\times$ \\
    128 & 256 / 1103 & 146 / 375 & 226 / 792 & 1.55$\times$ / 2.11$\times$\\
    \Xhline{1.0pt}
\end{tabular}
\vspace{-1em}
\caption{Comparing the training and inference speeds (images per second) of SwinV2-MoE using Fairseq and \tutel{}.}
\vspace{-1em}
\label{tab:speed_swin_moe}
\end{table}

\subsubsection{Experiment Results}

\paragraph{Speed Comparison.} \cref{tab:speed_swin_moe} compares the training and inference speeds of SwinV2-MoE using Fairseq and \tutel{}. For all GPU numbers, from 8 to 128 (1 expert per GPU), \tutel{} is significantly faster than Fairseq in both training and inference. Speedup of each iteration is $1.14\times \sim \bm{1.55\times}$ and $1.95\times \sim \bm{2.11\times}$ in training and inference, respectively.

\begin{table}[t]
    \centering
    \small
    \addtolength{\tabcolsep}{-3pt}
    \begin{tabular}{cccccc}
    \Xhline{1.0pt}
    Method & \makecell{IN-22K \\ acc@1} & \makecell{IN-1K/ft\\acc@1} & \makecell{IN-1K/5-shot\\acc@1} &  \makecell{COCO (AP)\\box / mask} \\
    \hline
    SwinV2-B & 37.2 & 85.1 & 75.9 & 53.0 / 45.8\\
    SwinV2-MoE-B & 38.5 & 85.5 & 77.9 & 53.4 / 46.2\\
    \Xhline{1.0pt}
    \end{tabular}
\vspace{-1em}
    \caption{Comparing the pre-training and fine-tuning accuracy between the sparse SwinV2-MoE-B model and its dense counterpart SwinV2-B.}
    \vspace{-1em}
    \label{tab:acc_swin_moe}
\end{table}

\paragraph{Accuracy Comparison.} We report the results of SwinV2-MoE-B on both pre-training and down-stream tasks, compared to the counterpart dense models, as shown in~\cref{tab:acc_swin_moe}. SwinV2-MoE-B achieves a top-1 accuracy of 38.5\% on the ImageNet-22K pre-training task, which is +1.3\% higher than the counterpart dense model. It also achieves higher accuracy on down-stream tasks: 85.5\% top-1 accuracy on ImageNet-1K image classification, 77.9\% top-1 accuracy on 5-shot ImageNet-1K classification, and 53.4/46.2 box/mask AP on COCO object detection, which is +0.4\%, +2.0\%, and +0.4/+0.4 box/mask AP higher than that using dense modes, respectively. In particular, it is the first time that the sparse MoE model is applied and demonstrated beneficial on the important down-stream vision task of COCO object detection.

\section{Conclusion}



In this paper, we analyze the key \textit{dynamic} characteristics in MoE from system's perspectives.
We address consequent issues by designing an \textit{adaptive} system for MoE, \tutel{}, which we present in two major aspects: adaptive parallelism for optimal expert execution and adaptive pipelining for tackling inefficient and non-scalable dispatch/combine operations in MoE layers.
We evaluate \tutel{} in an Azure A100 cluster with 2,048 GPUs and show that it achieves up to $5.75\times$ speedup for a single MoE layer.
\tutel{} empowers both training and inference of real-world state-of-the-art deep learning models.
As an example, this paper introduces our practice that adopts \tutel{} for developing SwinV2-MoE, which shows effectiveness of MoE in computer vision tasks comparing against the counterpart dense model.

\section*{Acknowledgements}

\blue{We appreciate the feedback by our shepherd, Lianmin Zheng, as well as anonymous reviewers of MLSys'23.}

\label{lastbody}

\bibliography{references}
\bibliographystyle{mlsys2023}

\clearpage


\appendix
\section{Two-dimensional Hierarchical (2DH) \AtoA{}}
\label{apx:2dh}

This section describes 2DH \AtoA{}, a novel \AtoA{} algorithm proposed by \tutel{}.

\subsection{Motivation: Small Size of Message Transfer}
\begin{figure}[t]
    \centering
    \begin{subfigure}[t]{.48\linewidth}
        \includegraphics[width=\linewidth]{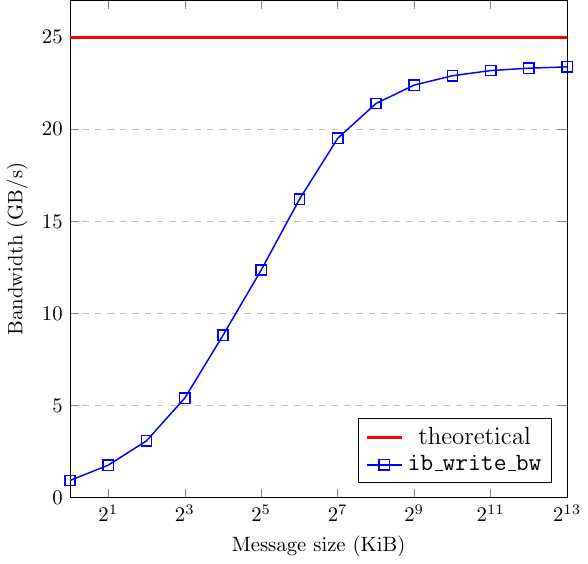}
        \caption{GPUDirect RDMA \texttt{ib\_write\_bw} (TX depth = 8) over HDR InfiniBand on two Azure NDv4 VMs~\cite{ndmv4}.}
        \label{fig:ibwrite}
    \end{subfigure}
    \hspace*{\fill}
    \begin{subfigure}[t]{.48\linewidth}
        \includegraphics[width=\linewidth]{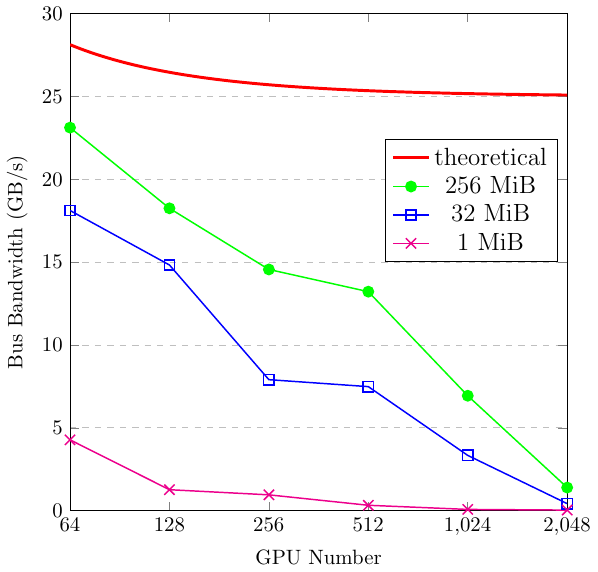}
        \caption{\AtoA{} bus bandwidth in nccl-tests scaling from 64-GPU to 2048-GPU.}
        \label{fig:alltoall_motivation}
    \end{subfigure}
    \caption{Under-utilized bandwidth for small messages.}
    \label{fig:msg_size}
\end{figure}

Most of popular DL frameworks~\cite{deepspeed,fairseq,sergeev2018horovod,pytorch} leverage point-to-point (P2P) APIs of NCCL~\cite{nccl-p2p},\footnote{Message Passing Interface (MPI)~\cite{snir1998mpi} also has developed various \AtoA{} algorithms~\cite{pjesivac2007towards,thakur1994all,bruck1997efficient}, but we only discuss NCCL in this work as it outperforms MPI in most DL scenarios. Note MPI mainly focuses on traditional HPC workloads where $S$ is typically much smaller than DL workloads.} the state-of-the-art GPU collective communication library, to implement \textit{Linear} \AtoA{} algorithm (see \cref{alg:p2p}).
It operates on $n$ GPUs, where each GPU splits its total $S$ bytes of data into $n$ chunks ($S/n$ bytes each) and performs P2P communication with all other GPUs.
The P2P chunk size $S/n$ transferred between any two GPUs will become smaller when we scale out (larger $n$), which is hard to saturate the high-speed links such as NVLink and HDR InfiniBand at a large scale (see \cref{fig:msg_size}). $S$ is fixed and only decided by the model itself.

\subsection{Approach and Challenges}
To achieve a high link bandwidth, our approach is aggregating multiple data chunks that are sent from multiple local GPUs to the same remote GPU.
This avoids sending multiple small messages over networking by merging small chunks into a single large chunk, which significantly improves the link bandwidth utilization.

Unfortunately, an efficient implementation of this approach on a large scale is challenging due to \emph{the overhead of aggregating small messages}.
Specifically, to aggregate chunks inside a node with $m$ local GPUs, all $m$ GPUs in the node need to exchange $\frac{S}{n} \times \frac{n}{m} = \frac{S}{m}$ chunks with each other. This is equivalent to performing $\frac{S}{n}$ size intra-node \AtoA{} $\frac{n}{m}$ times, as illustrated in~\cref{fig:2dh-a2a}, phase~1 of the na\"ive local aggregation \AtoA{}.
The latency of this intra-node \AtoA{} process is expected to be constant as chunk size $\frac{S}{m}$ does not rely on $n$, but unexpectedly, it actually increases as $n$ scales out due to $\frac{n}{m}$ times non-contiguous memory access on GPUs. For example, in phase 1 of the na\"ive local aggregation, intra-node GPUs exchange non-contiguous chunks twice with each other (\texttt{01} and \texttt{05}, \texttt{02} and \texttt{06}, etc.) that incurs $\mathcal{O}(\frac{n}{m})$ non-contiguous memory access on each GPU.
Specifically, when $S=128~\texttt{MiB}$ and $m=8$, we observe that intra-node \AtoA{} process takes $\sim 600 \mu s$ for $n=8$ and increases up to $\sim 5 ms$ for $n=2048$.

\subsection{Algorithm}
\begin{figure*}[h]
    \centering
    \includegraphics[width=\linewidth]{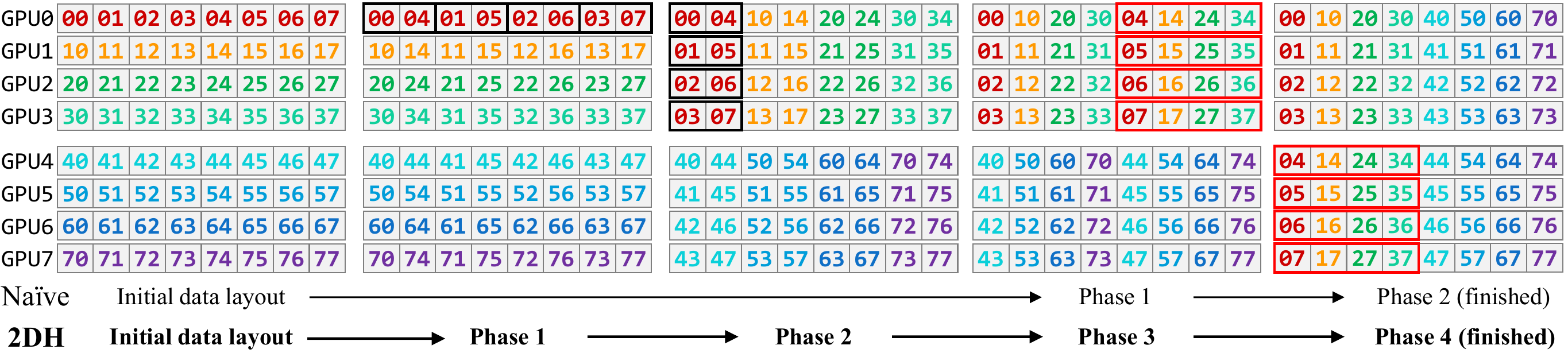}
    \caption{Example of data layouts in each phase of the na\"ive local aggregation \AtoA{} and two-dimensional hierarchical (2DH) \AtoA{}. In this example, there are two nodes that consist of GPU~0$\sim$3 and GPU~4$\sim$7, respectively.}
    \label{fig:2dh-a2a}
\end{figure*}

To avoid the slowdown due to non-contiguous memory access, 2DH \AtoA{} consists of additional phases that conduct efficient stride memory copies to align non-contiguous chunks into a contiguous address space. To be specific, \cref{fig:2dh-a2a} illustrates all phases of 2DH \AtoA{} in order. Instead of performing intra-node \AtoA{} from the beginning like the na\"ive local aggregation, we first align chunks that share the same local destination GPU via stride memory copies (phase~1) and then conduct intra-node \AtoA{} (phase~2). In the following phase, again, we align chunks that share the same remote destination GPU (phase~3) and then finally conduct inter-node \AtoA{} (phase~4).
By leveraging stride memory copies, 2DH \AtoA{} achieves a high memory bandwidth utilization, keeping a constant and low latency regardless of $n$ in the first three phases.
The benefit of 2DH \AtoA{} over existing algorithms increases as $S/n$ gets smaller (a smaller data size $S$ or a larger number of GPUs $n$).
Note that this is beneficial for rail-optimized InfiniBand networking as well since it avoids cross-rail communication.

\begin{algorithm}[t]
\caption{Linear \AtoA{} using Point-to-Point APIs}
\label{alg:p2p}
\begin{algorithmic}[1]
\PROCEDURE{All2All\_Linear}{\texttt{output, input}}
    \STATE \texttt{n $\gets$ ngpus, S $\gets$ sizeof input}
    \STATE \texttt{chunksize $\gets$ S / n}
    \FOR{\texttt{r = 0; r < n; r++}} 
        \STATE \texttt{loc $\gets$ r $\times$ chunksize, peer $\gets$ r}
        \STATE \texttt{ncclSend(input[loc], chunksize, peer)}
        \STATE \texttt{ncclRecv(output[loc], chunksize, peer)}
    \ENDFOR 
\ENDPROCEDURE
\end{algorithmic}
\end{algorithm}

\begin{figure*}[t]
    \centering
    \subfloat[\AtoA{} 1 MiB.]{\label{fig:2d1m}{\includegraphics[width=0.33\linewidth]{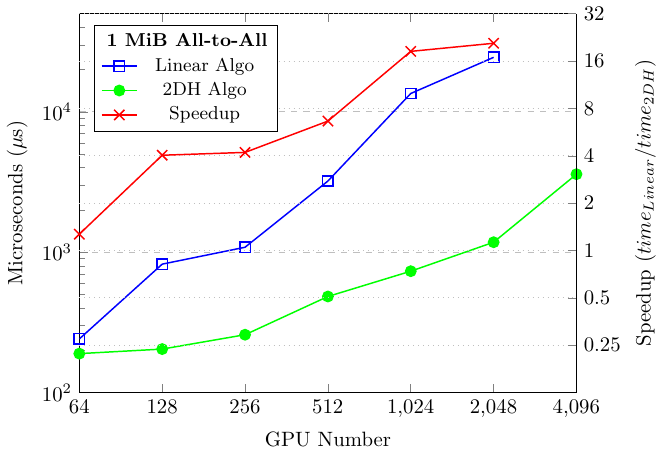}}}
    \
    \subfloat[\AtoA{} 32 MiB.]{\label{fig:2d32m}{\includegraphics[width=0.33\linewidth]{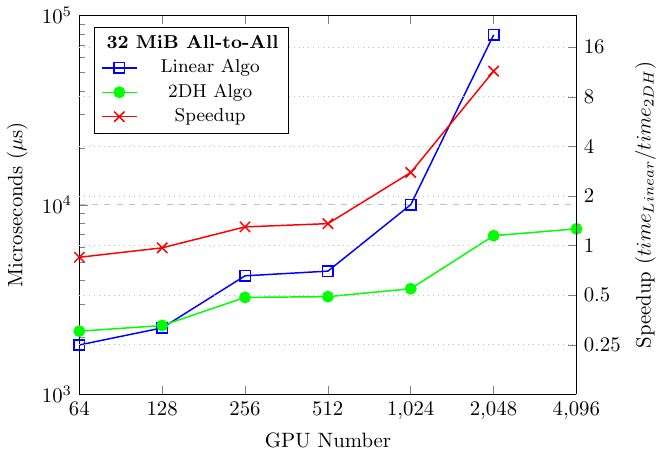}}}
    \
    \subfloat[\AtoA{} 256 MiB.]{\label{fig:2d256m}{\includegraphics[width=0.33\linewidth]{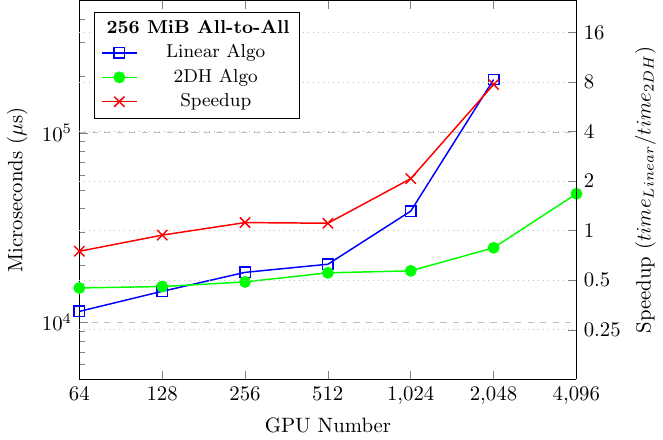}}}
    \caption{Comparison between linear and 2DH \AtoA{} algorithms with various sizes in NCCL.}
    \label{fig:2d}
\end{figure*}

\subsection{Optimization with \ifblind\else MSCCL\fi}

\paragraph{Implementation using NCCL APIs.}
We implement 2DH \AtoA{} algorithm using NCCL's \texttt{ncclSend} and \texttt{ncclRecv} APIs (see details in~\cref{alg:2dh}).
It consists of two steps. The first step corresponds to phase $1\sim3$ in~\cref{fig:2dh-a2a} and contains intra-node \AtoA{} communication and two stride memory copies, of which latencies only rely on $S$.
The second step corresponds to phase 4 in~\cref{fig:2dh-a2a}, which is inter-node \AtoA{} and its latency relies on $n/m$ instead of $n$ as local chunks are already merged.

\paragraph{Optimization via MSCCL.}
Implementation using NCCL APIs requires extra synchronization barriers between different phases in 2DH \AtoA{} and may cause throughput degradation.
In order to achieve better performance, we leverage \ifblind\else MSCCL \fi by describing the 2DH algorithm in a domain specific language (DSL) and optimizing with the compiler~\ifblind\else\cite{msccl}\fi. The custom compiler also leverages LL128 protocol~\cite{ll128} for \AtoA{}, which could achieve better efficiency than default NCCL-based implementation in low latency scenarios like small sizes \AtoA{}.

\paragraph{Extension.}
On existing GPU clusters, local GPU number $m$ is usually 8 or 16, which makes $\frac{n}{m}$ still large when scaling out \AtoA{} to hundreds of thousands (100 K) of GPUs \emph{at exascale}. The next generation NVSwitch~\cite{nvlink} enables up to 256 GPUs connected via high speed NVLink and makes it possible for 2DH \AtoA{} scaling out with $m=256$. For large-scale network topologies like dragonfly~\cite{kim2008technology}, 2DH \AtoA{} could be further adapted to 3D by splitting inter-node to intra-group and inter-group \AtoA{} according to the network hierarchy.

\begin{algorithm*}[t]
\caption{Two-Dimensional Hierarchical (2DH) \AtoA{}}
\label{alg:2dh}
\begin{algorithmic}[1]
\PROCEDURE{StrideMemcpy}{\texttt{output, input, chunksize, row, col}}
    \FOR{\texttt{i = 0; i < row $\times$ col; i++}}
        \STATE \texttt{j} $\gets$ \texttt{i \% row $\times$ col + i / col}
        \STATE \texttt{output[j $\times$ chunksize : (j+1) $\times$ chunksize]} $\gets$ \texttt{input[i $\times$ chunksize : (i+1) $\times$ chunksize]}
    \ENDFOR
\ENDPROCEDURE

\PROCEDURE{All2All\_2DH}{\texttt{output, input}}
    \STATE \texttt{// step 1: intra-node \AtoA{}}
    \STATE \texttt{strideMemcpy(buffer, input, chunksize, ngpus\_per\_node, nnodes)}
    \FOR{\texttt{g = 0; g < ngpus\_per\_node; g++}} 
        \STATE \texttt{loc $\gets$ g $\times$ nnodes $\times$ chunksize, peer $\gets$ g + node\_rank $\times$ ngpus\_per\_node}
        \STATE \texttt{ncclSend(buffer[loc], nnodes $\times$ chunksize, datatype, peer, comm)}
        \STATE \texttt{ncclRecv(output[loc], nnodes $\times$ chunksize, datatype, peer, comm)}
    \ENDFOR 
    \STATE \texttt{strideMemcpy(buffer, output, chunksize, nnodes, ngpus\_per\_node)}
    \STATE \texttt{// step 2: inter-node \AtoA{}}
    \FOR{\texttt{n = 0; n < nnodes; n++}} 
        \STATE \texttt{loc $\gets$ n $\times$ ngpus\_per\_node $\times$ chunksize, peer $\gets$ local\_rank + n $\times$ ngpus\_per\_node}
        \STATE \texttt{ncclSend(buffer[loc], ngpus\_per\_node $\times$ chunksize, datatype, peer, comm)}
        \STATE \texttt{ncclRecv(output[loc], ngpus\_per\_node $\times$ chunksize, datatype, peer, comm)}
    \ENDFOR 
\ENDPROCEDURE
\end{algorithmic}
\end{algorithm*}

\begin{figure*}[t]
    \centering
    \begin{subfigure}{.3\linewidth}
        \includegraphics[width=\linewidth]{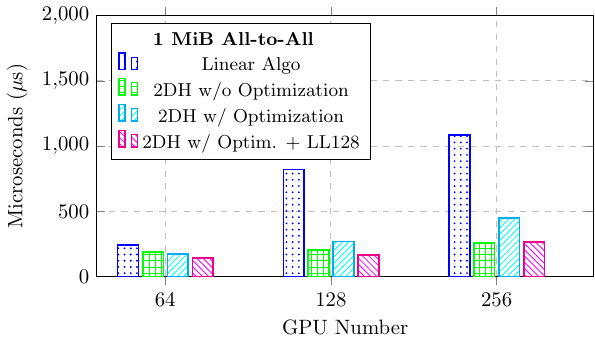}
        \caption{\AtoA{} 1 MiB}
        \label{fig:msccl1m}
    \end{subfigure}
    \hspace*{\fill}
    \begin{subfigure}{.3\linewidth}
        \includegraphics[width=\linewidth]{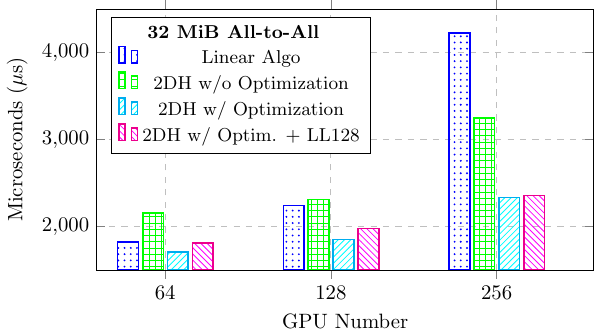}
        \caption{\AtoA{} 32 MiB}
        \label{fig:msccl32m}
    \end{subfigure}
    \hspace*{\fill}
    \begin{subfigure}{.3\linewidth}
        \includegraphics[width=\linewidth]{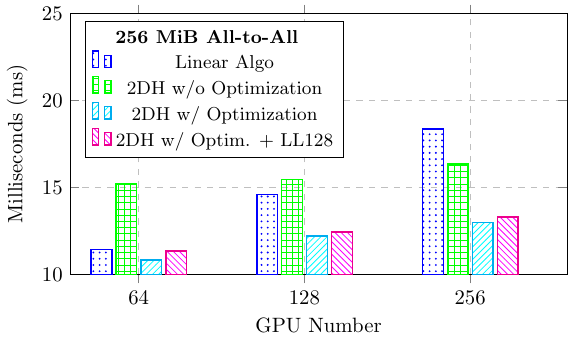}
        \caption{\AtoA{} 256 MiB}
        \label{fig:msccl256m}
    \end{subfigure}
    \caption{Comparison between NCCL and optimized implementation~\ifblind\else\cite{msccl}\fi~running 2DH \AtoA{} algorithm.}
    \label{fig:msccl}
\end{figure*}

\subsection{Evaluation}
We benchmark \texttt{alltoall\_perf} in nccl-tests~\cite{nccltests} to measure the performance and correctness of \AtoA{} operations.
Experiment setup is as described in \cref{sec:eval}.
The sizes of \AtoA{} start from 1 KiB and end at 16 GiB, with multiplication factor 2.
The tests are launched via OpenMPI with proper NUMA binding.
All of the \AtoA{} operations are out-of-place and correctness is also checked by nccl-tests.
We compare the latency of specific sizes we are interested in between different algorithms and different implementations.

To illustrate scalability of the proposed 2DH \AtoA{} algorithm, we compare it with the state-of-the-art NCCL \AtoA{} in the same cluster. \texttt{alltoall\_perf} in nccl-tests~\cite{nccltests} uses the linear \AtoA{} algorithm by default while we also implement the 2DH \AtoA{} algorithm in nccl-tests to replace the original one.
We scale the experiments from 64-GPU to 4096-GPU. As shown in~\cref{fig:2d}, the proposed 2DH algorithm could scale better with lower gradient than original linear algorithm. For small sizes (1 MiB), 2DH algorithm can achieve lower latency starting from small scales. For larger sizes (32 MiB and 256 MiB), 2DH algorithm has higher latency caused by extra data copies. While as the GPU number scales out, 2DH algorithm could perform better. Therefore, dynamic adaption between linear and 2DH algorithms is required. Besides, the 2DH algorithm can scale to 4096-GPU in our experiments while we didn't run NCCL's linear algorithm successfully in such large scale.

We also study the performance gain using the custom compiler~\ifblind\else\cite{msccl}\fi.
As illustrated in~\cref{fig:msccl}, the optimized implementation achieves better results than implementation using NCCL's APIs. For example, 256 MiB size on 64-GPU, 2DH algorithm in NCCL implementation has higher latency, but with the optimized implementation it could still outperform linear algorithm in NCCL.
Besides, LL128 protocol has lower latency for small sizes (1 MiB and 32 MiB) while default protocol performs better for large sizes (256 MiB). Therefore, dynamic adaption between different protocols is necessary with this optimization.

\section{SIMT-efficient Fast Encode and Decode}
\label{apx:fast-enc-dec}

\begin{figure}[t]
    \centering
    \subfloat[Dense implementation.]{\label{fig:dense-comp}{\includegraphics[width=\linewidth]{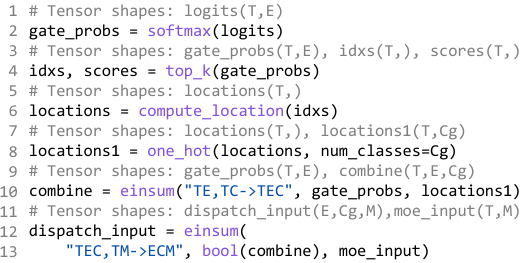}}}
    \vspace{1em}
    \subfloat[Sparse implementation.]{\label{fig:sparse-comp}{\includegraphics[width=\linewidth]{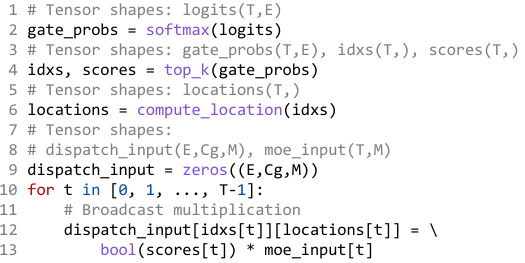}}}
    \caption{Comparison between dense and sparse implementations of generating \AtoA{} dispatch input (\texttt{dispatch\_input}) out of an MoE layer input (\texttt{moe\_input}) and a gate function output (\texttt{logits}).}
    \label{fig:dense-sparse-comp}
\end{figure}

\tutel{} implements sophisticated optimizations for the \textit{encode} (generating \AtoA{} inputs out of MoE layer inputs during MoE dispatch) and \textit{decode} (generating MoE layer outputs out of \AtoA{} outputs during MoE combine) stages of an MoE layer. Existing implementations of encode and decode need einsum operations with a large time complexity, as described by GShard~\cite{gshard} and implemented in Fairseq~\cite{fairseq}. For instance, \cref{fig:dense-comp} shows the most heavy-weighted part of the encode implementation (decode is similar as encode since it is a reverse operation of encode). We observe that this implementation is unnecessarily dense as it contains a lot of zero multiplications and additions. \tutel{} addresses this by a sparse implementation as shown in \cref{fig:sparse-comp}. Given that $T$ is the number of input tokens per expert, while the time complexity of the dense version is $O(T\cdot E\cdot C_g\cdot D)$, the one of the sparse version is only $O(T\cdot k\cdot D)$, where $T\cdot k = E\cdot C_g$ in most cases. 
This indicates that the sparse version has only $1/T$ of time complexity than the dense version.

\begin{figure}[t]
  \centering
    \includegraphics[width=\linewidth]{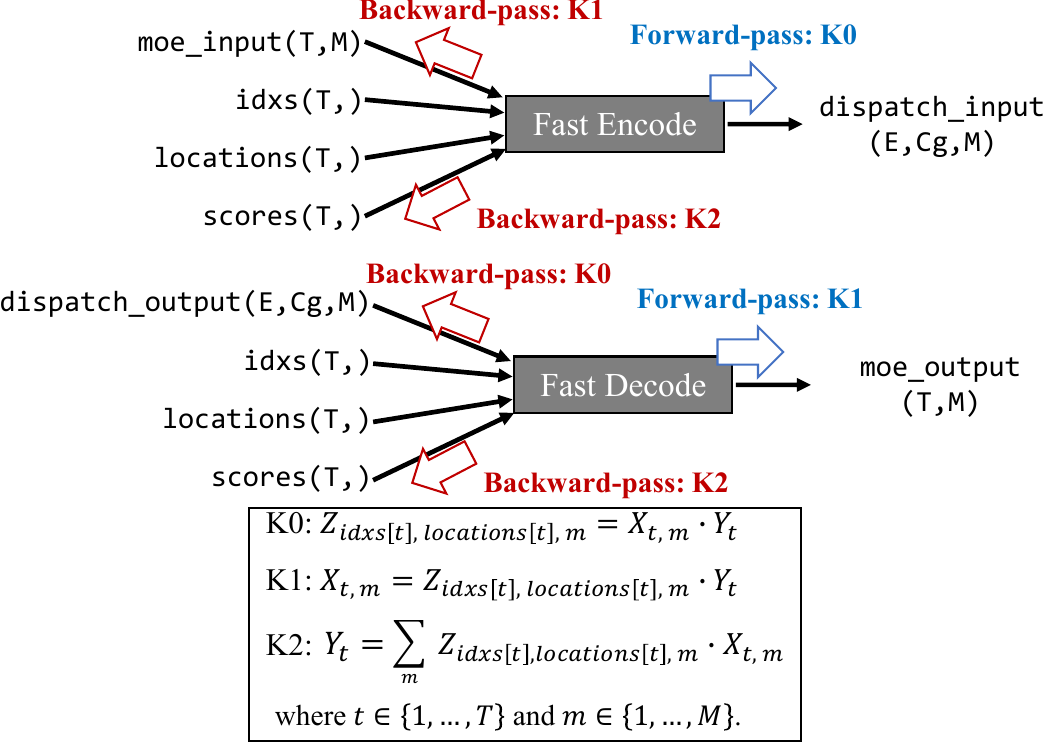}
  \caption{Forward- and backward-pass computations of fast encode and fast decode operators. Parentheses refer to tensor shapes. The tensor shapes of \texttt{X}, \texttt{Y}, and \texttt{Z} are $(T,D)$, $(T,)$, and $(E,C_g, D)$, respectively. \texttt{idxs} and \texttt{locations} have no backward-pass computation as they are not trainable inputs.}
  \label{fig:fast-enc-dec}
\end{figure}

Unfortunately, it is challenging to implement efficient GPU kernels for the sparse implementation. While the dense computation can be dramatically accelerated by matrix multiplication accelerators (e.g., Tensor Cores), the sparse computation cannot leverage those accelerators efficiently.\footnote{Even the sparsity support by the latest hardware (e.g., 3rd-generation Tensor Cores) cannot work efficiently as it only supports fine-grained sparsity, while our sparse computation belongs to coarse-grained sparsity~\cite{nvidia-ampere}.}

To tackle this issue, we implement differentiable fast encode and decode operators based on three specially designed GPU kernels: K0, K1, and K2, as illustrated in \cref{fig:fast-enc-dec}. \tutel{} accelerates these kernels by always assigning different indices of dimension $T$ to different thread arrays (or \textit{warps}), which ensures
computation for a single token along dimension $M$ is SIMT-efficient.
By this approach, our sparse computation can actually leverage various optimizations that are applicable only for dense computation, such as warp shuffling, Blelloch scan algorithm, and element vectorization for low-precision computation (e.g., leveraging half2 types for half-precision computation).
Aggregating all the kernel optimizations, \tutel{} extremely minimizes the latency of encode and decode as shown in~\cref{fig:comp-breakdown}. It greatly saves GPU memory as well. As shown in~\cref{tab:memory-apx}, in most cases, it achieves $20\% \sim 90\%$ memory saving. \tutel{} exposes two interfaces for these optimized computations: \texttt{moe.fast\_encode} used by MoE dispatch and \texttt{moe.fast\_decode} used by MoE combine.

\begin{table}[t]
\centering\small
\begin{tabular}{c|c|c}
    \Xhline{1.0pt}
    tokens/step & Fairseq MoE (GiB) & \tutel{} MoE (GiB) \\\hline
    4,096 & 3.7 & 2.9 ~~ (-21.6\%) \\\hline
    8,192 & 6.2 & 3.2 ~~ (-48.4\%)  \\\hline
    16,384 & 16.3 & 4.0 ~~ (-75.5\%) \\\hline
    32,768 & 57.9 & 5.7 ~~ (-90.2\%) \\\Xhline{1.0pt}
\end{tabular}
\vspace{-0.1in}
\caption{GPU memory cost for single MoE layer. (Static Settings: $D = H = 4096$, top-k = 2, $E_g$ = 2)}
\label{tab:memory-apx}
\end{table}

\section{More Results on SwinV2-MoE}

\subsection{How to do fine-tuning on COCO object detection?}

Previous MoE models on computer vision only perform experiments using image classification tasks~\cite{v-moe}. It is unclear whether the sparse MoE models perform well on down-stream computer vision tasks as well such as COCO object detection.

As shown in~\cref{tab:compare_coco}, direct fine-tuning will result in poor performance, with -1.7/-1.4 box/mask AP drops compared to the dense counterparts. We find that fixing all MoE layers in fine-tuning can alleviate the degradation problem, and we obtain +0.4/+0.4 box/mask AP improvements by this strategy.

Also note it is the first time that a sparse MoE model is applicable and superior on the important computer vision tasks of COCO object detection. We hope \tutel{} to empower more down-stream AI tasks.

\begin{table}[t]
    \centering
    \small
    \addtolength{\tabcolsep}{-2.2pt}
    \begin{tabular}{ccccccc}
    \Xhline{1.0pt}
    Method & $E$ & $k$ & $f$ & MoE & AP$^{box}$ &AP$^{mask}$\\
    \hline
    SwinV2-B & - &  - & - & - & 53.0 & 	45.8 \\
    \hline
    SwinV2-MoE-B & 32 & 1 & 1.25 & tuned &51.3 (-1.7) & 44.4 (-1.4) \\
    SwinV2-MoE-B & 32 & 1 & 1.25 & fixed &53.4 (+0.4)  & 46.2 (+0.4)\\
    \Xhline{1.0pt}
    \end{tabular}
    \caption{The results on COCO object detection. ``fixed" MoE indicates that the MoE layers are fixed in fine-tuning. }
    \label{tab:compare_coco}
\end{table}

\begin{table*}[t]
    \centering
    \small
    \addtolength{\tabcolsep}{-2.5pt}
    \begin{tabular}{cccccccccccccc}
    \Xhline{1.0pt}
    Method & $E$ & $k$ & $f$ & \#$param$ & \#$param_{act}$ & \footnotesize{GFLOPs} & \makecell{Train\\speed} & \makecell{Inference\\speed} & \makecell{IN-22K\\acc@1} & \makecell{IN-22K\\ train loss} & \makecell{IN-1K/ft\\acc@1} & \makecell{IN-1K/5-shot\\acc@1} \\
    \hline
    SwinV2-S & - & - & - & 65.8M & 65.8M & 6.76 & 350 & 1604 & 35.5 & 5.017 &83.5  & 70.3\\
    SwinV2-MoE-S & 8 & 1 & 1.0 & 173.3M & 65.8M & 6.76 & 292 & 1150 & 36.8 (+1.3) & 4.862  &84.5 (+1.0)   & 75.2 (+4.9)\\
    SwinV2-MoE-S & 16 & 1 & 1.0 & 296.1M & 65.8M &6.76 & 295 & 1153 & 37.5 (+2.0) & 4.749  & 84.9 (+1.4)  & 76.5 (+6.2)\\
    SwinV2-MoE-S & 32 & 1 & 1.0 & 541.8M & 65.8M &6.76 & 295 & 1159 & 37.4 (+1.9) & 4.721  & 84.7 (+1.2) & 75.9 (+5.6)\\
    SwinV2-MoE-S & 64 & 1 & 1.0 & 1033M & 65.8M & 6.76  & 288 & 1083 & 37.8 (+2.3) & 4.669 & 84.7 (+1.2) & 75.7 (+5.4)\\
    SwinV2-MoE-S & 128 & 1 & 1.0 & 2016M & 65.8M & 6.76  & 273 & 1027 & 37.4 (+1.9) & 4.744 & 84.5 (+1.0)  & 75.4 (+5.1)\\
    \hline
    SwinV2-B & - & - & - &109.3M & 109.3M & 11.78 & 288 & 1195 & 37.2 & 4.771 & 85.1 & 75.9 \\
    SwinV2-MoE-B & 8 & 1 & 1.0 &300.3M & 109.3M &11.78  & 247 & 893 &38.1 (+0.9)&4.690& 85.3 (+0.2)  & 77.2 (+1.3)\\
    SwinV2-MoE-B & 16 & 1 & 1.0 & 518.7M & 109.3M & 11.78& 246 & 889  & 38.6 (+1.4)& 4.596& 85.5 (+0.4)  & 78.2 (+2.3)\\
    SwinV2-MoE-B & 32 & 1 & 1.0 & 955.3M & 109.3M & 11.78 & 249 & 892 & 38.5 (+1.3)&4.568&85.5 (+0.4)  & 77.9 (+2.0)\\
    SwinV2-MoE-B & 32 & 2 & 1.0 & 955.3M & 136.6M & 11.78 & 206 & 679 & 38.6 (+1.4) & 4.506 & 85.5 (+0.4) & 78.7 (+2.8) \\
    SwinV2-MoE-B & 32 & 2 & 0.625 & 955.3M & 136.6M & 12.54 & 227 & 785 & 38.3 (+1.1) & 4.621 &  85.2 (+0.1) & 77.5 (+1.6)\\
    \Xhline{1.0pt}
    \end{tabular}
    \caption{Comparison of SwinV2-MoE models and the dense counterparts~\cite{liu2021swinv2}. The sparse MoE model is obtained by replacing the FFN of every other layer with an MoE layer. $E$ denotes the number of experts in the MoE layer. $k$ denotes the number of selected experts per token. $f$ denotes the capacity factor. The “Train speed” and “Inference speed" are measured by images per second during training and inference. All models are trained on the ImageNet-22K dataset with an input resolution of $192\times192$. We report the top-1 accuracy and final training loss on ImageNet-22K classification (IN-22K), the fine-tuning top-1 accuracy on ImageNet-1K classification (IN-1K/ft) and the 5-shot linear evaluation top-1 accuracy on ImageNet-1K classification (IN-1K/5-shot). Also note that \tutel{} supports multiple GPUs to share one expert, which empowers us to leverage 32 GPUs for the experiments with expert number as 8 and 16. }
    \label{tab:swin_e2e}
\end{table*}

\subsection{Ablation Study}

\paragraph{Ablation on Number of Experts.}
\cref{tab:swin_e2e} ablates the effect of expert number, using different model sizes (SwinV2-S and SwinV2-B) and a variety of vision tasks. It can be seen that 32 and 64 perform the best, which is consistent with that in previous works~\cite{v-moe,glam}.

\begin{figure}[t]
    \centering
    \includegraphics[width=0.8\linewidth]{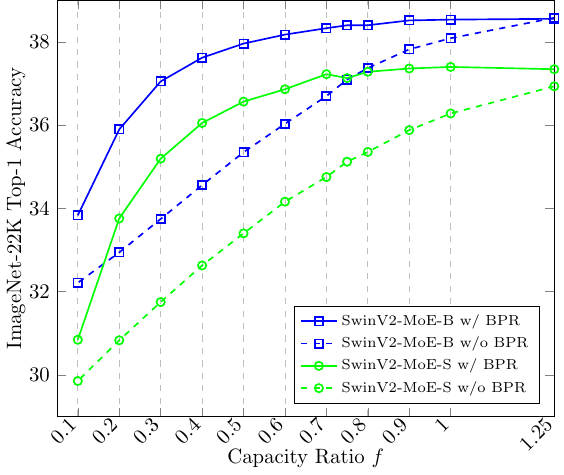}
    \caption{ImageNet-22K top-1 accuracy w.r.t inference capacity factor. ``w/ BPR" indicates training with batch prioritized routing while ``w/o BPR" not. All models are trained on the ImageNet-22K dataset with $E=32$, $k=1$, $f=1.25$ and an input resolution of $192\times192$ for 90 epochs.} 
    \label{fig:compare_bpr}
\end{figure}

\paragraph{Comparison of Routing Algorithms and Capacity Factors.}
\cref{fig:compare_bpr} compares the routing methods with and without batch prioritized routing (BPR)~\cite{v-moe}. It shows that the BPR approach is crucial for computer vision MoE models, especially at lower capacity factor values. These results are consistent with reported in~\cite{v-moe}.

\cref{tab:compare_topk_cf} ablates the performance of SwinV2-MoE model given different $k$ and capacity factor $f$. It is observed that top-1 router has a better speed-accuracy trade-off. We use default hyper-parameters of $k=1$ and $f=1.0$.
 
\subsection{A New Cosine Router Supported in \tutel{}}

With \tutel{}, we provide more MoE baselines to enrich the algorithm choices and to exemplify how to leverage this framework for algorithmic innovation. One attempt is a new cosine router that hopes to improve numerical stability with increased model size, inspired by~\cite{liu2021swinv2}:
\begin{equation}
\label{eq.cosine}
    P = \text{Softmax}(\frac{W\mathbf{x} \cdot M}{\left \| W\mathbf{x} \right \|  \left \| M \right \|} / \tau),
\end{equation}
where $W \in \mathbb{R}^{D \times C}$ is a linear layer used to project the input token feature $x \in \mathbb{R}^{C\times 1}$ to dimension $D$ (256 by default); $M \in \mathbb{R}^{E \times D}$ is a parametric matrix, with each column representing each expert; $\tau$ is a learnable temperature that is set lowest 0.01 to avoid temperatures being too small; $P$ denotes the routing scores for selecting experts. 

Our preliminary experiments in~\cref{tab:cosine_router} show that when using 32 experts, the cosine router is as accurate in image classification as a common linear router. Although it is not superior in image classification at the moment, we still encourage \tutel{} users to try this option in their problems, because: 1) its normalization effect on input may lead to more stable routing when the amplitude or dimension of the input feature is scaled; 2) There is a concurrent work showing that the cosine router is more accurate in cross-lingual language tasks~\cite{xmoe22}.

\begin{table}[t]
    \centering
    \small
    \addtolength{\tabcolsep}{-3pt}
    \begin{tabular}{ccccccc}
    \Xhline{1.0pt}
    Method & $k$ & Train-$f$ & Infer-$f$ &\makecell{Infer\\\footnotesize{GFLOPs}} &  \makecell{Infer\\speed} & \makecell{IN-22K\\acc@1}\\
    \hline
    \footnotesize{SwinV2-B} & - &  - & - & 11.78 & 1195 & 37.2 \\
    \hline
    \footnotesize{SwinV2-MoE-B} &  1 & 1.0 & 1.25 & 12.54 & 839 & 38.6 (+1.4)\\
    \footnotesize{SwinV2-MoE-B} &  1 & 1.0 & 1.0  & 11.78 & 892 & 38.5 (+1.3)\\
    \footnotesize{SwinV2-MoE-B} &  1 & 1.0 & 0.625 & 10.65  & 976 & 38.2 (+1.0)\\
    \footnotesize{SwinV2-MoE-B} &  1 & 1.0 & 0.5 & 10.27 & 1001 & 38.0 (+0.8) \\
    \hline
    \footnotesize{SwinV2-MoE-B} &  2 & 1.0 & 1.25 & 16.31 & 621 & 38.7 (+1.5)\\
    \footnotesize{SwinV2-MoE-B} &  2 & 1.0 & 1.0  & 14.80 & 679 & 38.6 (+1.4)\\
    \footnotesize{SwinV2-MoE-B} &  2 & 1.0 & 0.625 &12.54 &785 & 38.4 (+1.2) \\
    \footnotesize{SwinV2-MoE-B} &  2 & 1.0 & 0.5 &11.78  & 826 & 38.3 (+1.1) \\
    \hline
    \footnotesize{SwinV2-MoE-B} &  2 & 0.625 & 0.625 & 12.54 & 785& 38.3 (+1.1)\\
    \footnotesize{SwinV2-MoE-B} &  2 & 0.625 & 0.5 & 11.78 & 826& 38.3 (+1.1) \\
    \Xhline{1.0pt}
    \end{tabular}
    \caption{Ablations of top-$k$ and capacity factors $f$. ``Train-$f$" and ``Infer-$f$" indicates the capacity factor during training and inference. ``Infer GFLOPs" and ``Infer speed" indicates the GFLOPs and real speed (images/second) during inference. }
    \label{tab:compare_topk_cf}
\end{table}

\begin{table}[t]
    \centering
    \small
    \addtolength{\tabcolsep}{-3pt}
    \begin{tabular}{ccccc}
    \Xhline{1.0pt}
    Method & Router & \makecell{IN-22K \\ acc@1} & \makecell{IN-1K/ft\\acc@1} & \makecell{IN-1K/5-shot\\acc@1} \\
    \hline
    SwinV2-S & - & 35.5 & 83.5 & 70.3\\
    SwinV2-MoE-S & Linear & 37.4 (+1.9) & 84.7 (+1.2) & 75.9 (+5.6)\\
    SwinV2-MoE-S & Cosine & 37.1 (+1.6) & 84.3 (+0.8) & 75.2 (+4.9)\\
    \hline
    SwinV2-B & - & 37.2 & 85.1 & 75.9 \\
    SwinV2-MoE-B & Linear & 38.5 (+1.3)& 85.5 (+0.4)& 77.9 (+2.0)\\
    SwinV2-MoE-B & Cosine & 38.5 (+1.3)& 85.3 (+0.2)& 77.3 (+1.4)\\
    \hline
    \Xhline{1.0pt}
    \end{tabular}
    \caption{Comparison between the linear router and cosine router ($E=32$, $k=1$, $f=1.25$).} 
    \label{tab:cosine_router}
\end{table}


\label{lastpage}
\end{document}